\documentclass[trackchanges, twocolumn]{aastex701}

\usepackage{multirow}
\usepackage{soulutf8} % Para hl{} funcionar

\begin{document}

\title{The Triple System V1371\,Tau: An Eclipsing Binary with an Outer Be Star}

\author[orcid=0000-0002-8225-7699]{Danilo F. Rocha}
\affiliation{Observat\'orio Nacional, R. Gen. Jos\'e Cristino, 77 - Vasco da Gama, Rio de Janeiro - RJ, 20921-400, Brazil.}
\affiliation{LIRA, Paris Observatory, PSL University, CNRS, Sorbonne University, Universit\'e Paris Cit\'e, Cergy CY Paris University, 5 place Jules Janssen, F-92195 Meudon, France.}
\affiliation{Laboratório Nacional de Astrofísica, Rua Estados Unidos 154, 37504-364 Itajubá, MG, Brazil.}
\email[show]{danilo-fr@live.com, danilorocha@on.br}

\author[orcid=0000-0001-5589-9015]{Marcelo Emilio}
\affiliation{Universidade Estadual de Ponta Grossa, Av. Carlos Cavalcanti, 4748, 84030-900 - Ponta Grossa, PR, Brazil.}
\affiliation{Observat\'orio Nacional, R. Gen. Jos\'e Cristino, 77 - Vasco da Gama, Rio de Janeiro - RJ, 20921-400, Brazil.}
\email{marcelo_emilio@yahoo.com}

\author[orcid=0000-0002-2919-6786]{Jonathan Labadie-Bartz}
\affiliation{LIRA, Paris Observatory, PSL University, CNRS, Sorbonne University, Universit\'e Paris Cit\'e, Cergy CY Paris University, 5 place Jules Janssen, F-92195 Meudon, France.}
\affiliation{DTU Space, Technical University of Denmark, Elektrovej 327, Kgs., Lyngby 2800, Denmark}
\email{jbartz@udel.edu}

\author[orcid=0000-0003-1978-9809]{Coralie Neiner}
\affiliation{LIRA, Paris Observatory, PSL University, CNRS, Sorbonne University, Universit\'e Paris Cit\'e, Cergy CY Paris University, 5 place Jules Janssen, F-92195 Meudon, France.}
\email{coralie.neiner@obspm.fr}

\author[orcid=0000-0002-9552-7010]{Julia Bodensteiner}
\affiliation{ESO - European Southern Observatory, Karl-Schwarzschild-Strasse 2, 85748 Garching bei M\"unchen, Germany.}
\email{j.bodensteiner@uva.nl}

\author[orcid=0000-0003-0642-8107]{Tomer Shenar}
\affiliation{The School of Physics and Astronomy, Tel Aviv University, Tel Aviv 6997801, Israel.}
\email{tshenar@tau.ac.il}

\author[orcid=0000-0001-9200-3441]{Laerte Andrade}
\affiliation{Laboratório Nacional de Astrofísica, Rua Estados Unidos 154, 37504-364 Itajubá, MG, Brazil.}
\email{landrade.1138@gmail.com}

\author[orcid=0000-0001-6566-7568]{Michael Abdul-Masih}
\affiliation{Instituto de Astrof\'isica de Canarias, C. V\'ia L\'actea, s/n, 38205 La Laguna, Santa Cruz de Tenerife, Spain.}
\affiliation{Universidad de La Laguna, Dpto. Astrof\'isica, Av.\ Astrof\'sico Francisco S\'anchez, 38206 La Laguna, Santa Cruz de Tenerife, Spain.}
\email{michael.abdulmasih@gmail.com}

\author[orcid=0000-0002-0284-0578]{Felipe Navarete}
\affiliation{Laboratório Nacional de Astrofísica, Rua Estados Unidos 154, 37504-364 Itajubá, MG, Brazil.}
\affiliation{SOAR Telescope/NSF's NOIRLab, Avda Juan Cisternas 1500, 1700000, La Serena, Chile.}
\email{fnavarete@lna.br}

\author[orcid=0000-0003-0468-3964]{Alessandro Melo}
\affiliation{Observat\'orio Nacional, R. Gen. Jos\'e Cristino, 77 - Vasco da Gama, Rio de Janeiro - RJ, 20921-400, Brazil.}
\email{alessandromelo310@gmail.com}

\author[orcid=0000-0003-0079-3912]{Eduardo Janot-Pacheco}
\affiliation{Instituto de Astronomia, Geof\'isica e Ci\^encias Atmosf\'ericas, Universidade de S\~ao Paulo, 05508-090, S\~ao Paulo, SP, Brazil.}
\email{eduardo.janot@iag.usp.br}

\author[orcid=0009-0003-6830-8044]{Romualdo Eleutério}
\affiliation{Observat\'orio Nacional, R. Gen. Jos\'e Cristino, 77 - Vasco da Gama, Rio de Janeiro - RJ, 20921-400, Brazil.}
\email{resj@discente.ifpe.edu.br}

\author[orcid=0000-0003-1669-6869]{Alan W. Pereira}
\affiliation{Universidade Estadual de Ponta Grossa, Av. Carlos Cavalcanti, 4748, 84030-900 - Ponta Grossa, PR, Brazil.}
\affiliation{Observat\'orio Nacional, R. Gen. Jos\'e Cristino, 77 - Vasco da Gama, Rio de Janeiro - RJ, 20921-400, Brazil.}
\email{alanwpereira@gmail.com}

\begin{abstract}

Although triple systems are common, their orbital dynamics and stellar evolution remain poorly understood. We investigated the V1371\,Tau system using TESS photometry, multi-epoch spectroscopy, and recent interferometric data, confirming it as a rare triple system consisting of a eclipsing binary orbited by a classical Be star, with a spectral classification of (B1V\,+\,B0V)\,+\,B0Ve. The eclipsing binary exhibits an orbital period of $\approx$\,34\,days, and the Be star orbits the inner pair on a timescale of a few years. Weak H$\alpha$ emission lines suggest the presence of a Keplerian disk with variability on a timescale of months around the Be star, and nearly constant V/R ratio no detectable asymmetry variations. Besides the eclipses, frequencies at 0.24 and 0.26 c/d dominate the photometric variability. Higher-frequency signals are present which appear associated with non-radial pulsation. The eclipsing pair ($i\,\approx\,90^\circ$) shows projected rotational velocities of 160 and 200\,km\,s$^{-1}$. The Be star’s measured $v\,\sin\,i \approx 250$\,km\,s$^{-1}$ implies a critical rotation fraction between 0.44 and 0.76 for plausible inclinations, significantly faster than the eclipsing components. The shallower eclipses in the KELT data compared to TESS suggest a variation in orbital inclination, possibly induced by Kozai–Lidov cycles from the outer Be star. The evolution analysis suggests that all components are massive main sequence stars, with the secondary star in the eclipsing binary being overluminous. This study emphasizes the complexity of triple systems with Be stars and provides a basis for future research on their formation, evolution, and dynamics.
 %needs 250 Words

\end{abstract}



%----------------------------------------------------
\section{Introduction} \label{sec:intro}

Classical Be stars are distinguished by prominent Balmer emission lines, which indicate the presence of circumstellar disks formed through decretion processes \citep{rivinius2013classical}. These emission features serve as evidence of material expelled from the stellar surface. Be stars are also notable for their rapid rotation, often approaching the critical break-up velocity. This extreme rotation, in combination with Non-Radial Pulsations (NRPs), is thought to drive episodic mass ejections that feed the formation and maintenance of the surrounding disks \citep{neiner2020transport}.

The study of Be stars in binary systems is of particular astrophysical interest, as these systems allow precise determination of fundamental parameters such as mass, radius, temperature, orbital separation, and eccentricity \citep{sowell2011absolute, Almeida15, poro2021new}, offering valuable insights into their individual and coupled evolution \citep{torres2010accurate, mahy2020tarantula}. Moreover, binary interactions involve additional complexity, with tidal forces contributing to the structure of those disks and the star's rotational dynamics. Binary systems play a crucial role in shaping stellar evolution \citep{sana2012binary}. When extended to multiple systems, such an analysis can significantly complicate our current insights into stellar evolution and often renders traditional models inadequate \citep{toonen2020evolution, kummer2023main}.
 
Hierarchical multiple systems, especially with triple stars, become increasingly common with higher stellar masses, with their frequency rising from roughly $\sim$10\% among low-mass stars to nearly $\sim$50\% among B-type stars \citep{duchene2013stellar, moe2017mind, maiz2019monos}. However, despite their prevalence, the evolution of such systems remains poorly understood due to the need for systematic studies. Notable examples include $\alpha$\,Cen, the nearest triple star system \citep{tokovinin2014binaries}, and $\nu$\,Gem, a hierarchical triple system where one component is a classical Be star \citep{klement2021nu, gardner2020armada}. These systems provide critical case studies for understanding the dynamics and evolution of complex stellar configurations, following pioneering work by \citet{fekel1981properties}.

In this context, we focus on V1371\,Tau\footnote{Coord. (J2000): R.A.\,=\,05h34m39.10s, Dec.\,=\,+28$^{\circ}$03$\arcmin$03.78$\farcs$} (TIC\,74386000, HD\,36665), a multiple star system comprising an inner eclipsing binary (EB) orbited by a classical Be star \citep{Rivinius2025}. Since it is a hierarchical system, the components will be identified as an external pair of binary system A\,+\,B, with A\,=\,a\,+\,b representing the eclipsing internal binary. This system is located within the supernova remnant (SNR) S147 (also known as Shajn 147 or Simeis 147) in a region rich with B-type stars \citep{minkowski_1958}. S147, known for its intricate filamentary structure spanning over 3 degrees of the sky \citep{dickel1969610}, is centered in the Taurus constellation. It represents a unique stage in stellar evolution, believed to have originated from a double supernova event \citep{gvaramadze2006supernova}. However, the exact age of S147 remains a topic of debate. At the same time, \citet{silk1973high} suggested an age of $9\,\times\,10^4$ years based on the high-velocity gas detected in the ultraviolet spectrum of V1371\,Tau, but at this age the SNR should already be very close to disappearing, since this estimate exceeds the typical age range of $10^4$ to $10^5$ years for evolved SNRs \citep{ferrand2012census} and contradicts the 30\,kyrs age inferred by \citet{kramer2003proper} from the pulsar PSR J0538 + 2817's motion within the remnant.

Hierarchical triple systems with eclipsing components and Be stars as outer components are rare. Understanding the role of gravitational interactions and mass transfer in these systems can provide critical insights into the formation and evolution of such stars. \citet{borkovits2025then} investigated 221 hierarchical triple systems candidates by combining Kepler and TESS data to analyze eclipse timing variations and uncover signs of precession and dynamical interactions. Among the systems studied, there is no mention of any Be-type stars or eclipsing sub-systems. Examples, such as $\nu$\,Gem \citep{klement2021nu}, HR\,2309, and possibly 141 Tau, stand out due to their unusual configuration, where the Be star occupies the outer position but is not the dominant source of light (For a discussion of such systems, see \citet{Rivinius2025}). Investigating such systems is crucial for exploring evolutionary pathways that may lead to the forming of Be stars in multiple systems.

The rapid rotation characteristic of Be stars, which brings them close to their critical velocity, can be explained by two main scenarios. In the case of a single star, the high rotation results from the angular momentum being transported from the contracting core to the surface during its evolution. As the core shrinks, it transfers part of its angular momentum to the outer layers, accelerating the star's surface rotation \citep{Granada2013}. Alternatively, this rapid rotation in binary systems may result from mass transfer between the stellar components. In such cases, the companion star, which evolves more rapidly, transfers material and angular momentum to the Be star, making it the system's highly rotating and visible component. 
Be stars rotate on average at about 65\% of their critical velocity, but they exhibit a wide range of true velocity ratios, typically between $V/V_c \approx 0.30$ and $0.95$ \citep{zorec2016critical}. However, such low rotation rates are likely not representative of actual Be stars, and may reflect uncertainties in the determination of fundamental parameters.

Studying the V1371\,Tau system allows one to explore these scenarios more thoroughly. A detailed analysis of V1371\,Tau may uncover evidence of past interactions that explain the rapid rotation of the Be star, such as mass transfer events or mergers. Moreover, this system offers a chance to identify possible evolutionary channels that lead to Be stars forming in hierarchical multiple systems. Therefore, this study can contribute to our understanding of the dynamic and evolutionary processes shaping these complex and rare systems.

The present manuscript is concerned with the detailed characterization of the physical and orbital properties of V1371\,Tau, aiming to understand the evolutionary history of the system. Previous studies often did not consider the multiplicity of this system, which may strongly affect our understanding of V1371\,Tau and the surrounding S147 remnant. Section\,\ref{sec:data} describes the photometric and spectroscopic data collection methods and reduction techniques employed. Section\,\ref{sec:binV1371Tau} presents an in-depth analysis of the eclipsing binary system within V1371\,Tau, utilizing state-of-the-art techniques to derive accurate orbital and physical parameters. In Section\,\ref{sec:beV1371Tau}, we explore the characteristics of the Be star component. Finally, Sections\,\ref{sec:V1371Tau} and \,\ref{sec:conclusion} synthesize our findings, discussing their implications for the broader context of hierarchical stellar systems and their evolution. This study opens new pathways for understanding the role of classical Be stars in complex stellar systems and their contribution to the galactic ecosystem.

\section{Data} \label{sec:data}

\subsection{Photometric Observations}

Launched in 2018, NASA's Transiting Exoplanet Survey Satellite (TESS) initially aimed to detect exoplanets via the transit method. However, its mission has since expanded to include the study of variable stars, pulsating stars such as Be stars, and binary systems \citep{southworth2022high, ramsay2022tess, burssens2020variability}. The satellite's continuous, high-cadence observations, free from atmospheric interference, make TESS an invaluable resource for stellar variability studies. This study used TESS data to investigate V1371\,Tau. TESS observed this target in two distinct periods: $\sim$76.5 days in 2021 (spanning Sectors 43-45) and $\sim$51.5 days in 2023 (Sectors 71-72). The observations were conducted with exposure times of 475 seconds in 2021 and 158 seconds in 2023, yielding a substantial dataset for our analysis. The light curves were extracted from these Full-Frame Images (FFIs) using the \texttt{Lightkurve} package in Python \citep{collaboration2018lightkurve}.

\begin{figure}
    \centering
    \includegraphics[width=\linewidth, trim={2.55cm 0.95cm 6.15cm 3.0cm},clip]{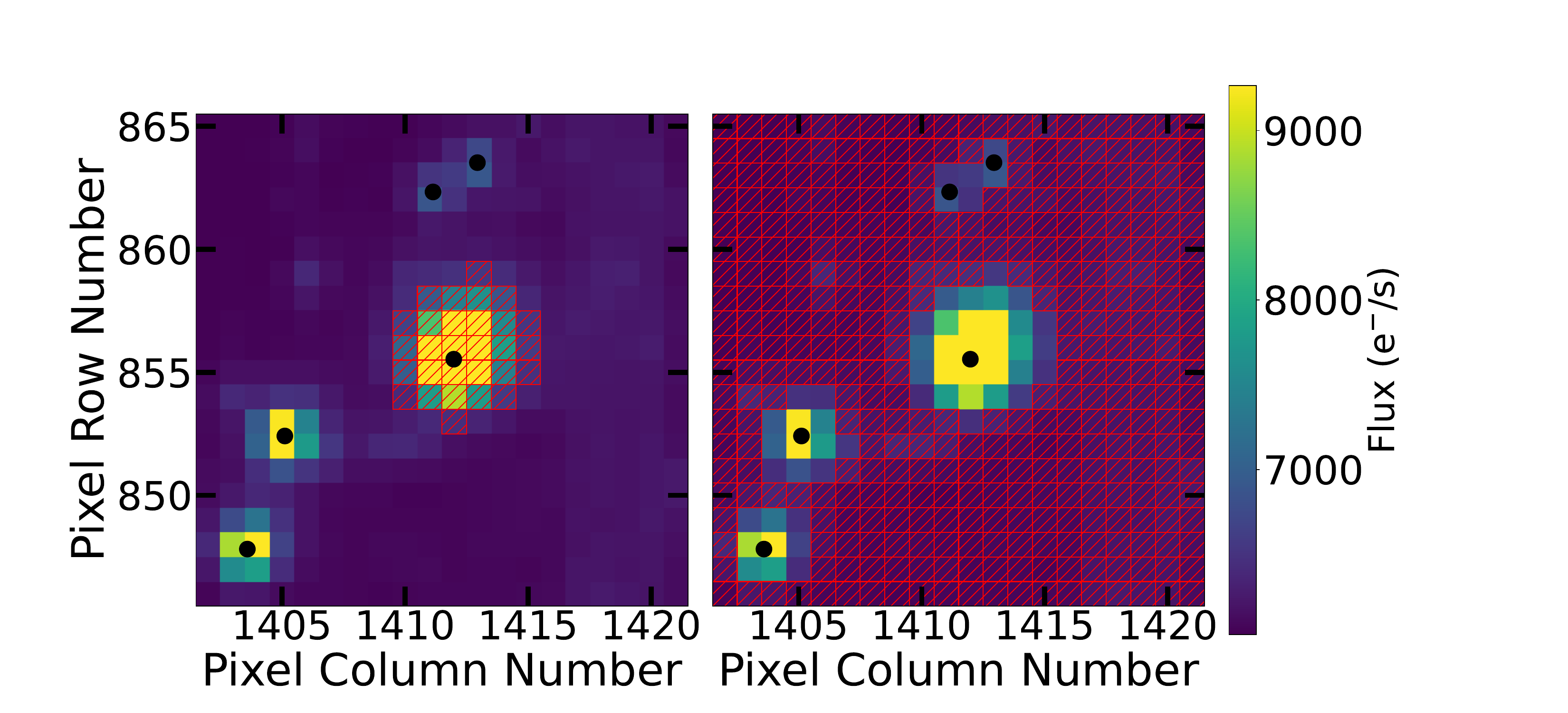} % Answer: [trim={left bottom right top},clip]
    \caption{The figure shows the aperture masks used to extract the flux of V1371\,Tau (left panel) and the background flux (right panel), as observed by TESS in Sector 43. Black dots indicate field sources from the Gaia catalog with magnitude $G \leq 13$.}
    \label{fig:tpf}
\end{figure}

\begin{figure*}
    \centering
    \includegraphics[width=\linewidth]{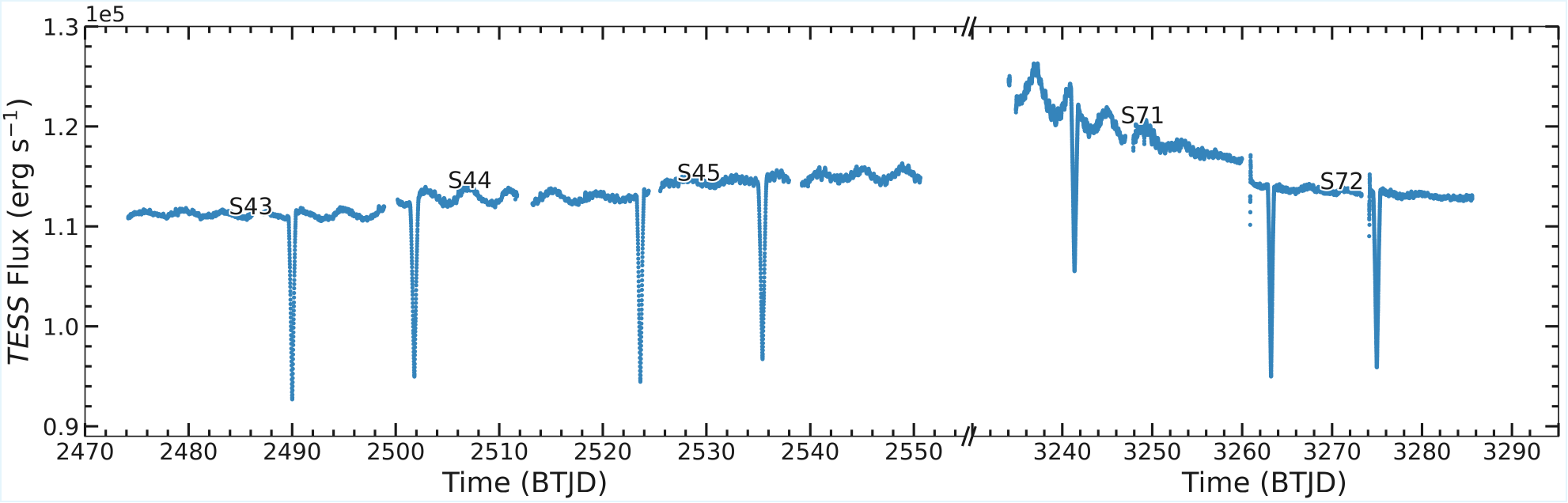} % Answer: [trim={left bottom right top},clip]
    \caption{Light curve of V1371\,Tau observed by TESS in flux at BTJD (BJD-2,457,000).}
    \label{fig:V1371Tau_lc}
\end{figure*}

To focus on the intrinsic light variations of our target star, V1371\,Tau, we employed the tool \texttt{TESScut} \citep{brasseur2019astrocut}. This procedure extracts and normalizes the target's flux, neighbouring celestial objects, and background noise. We defined a region of interest—a Target Pixel File (TPF)—with a square with 20 pixels on each side centred on the star's coordinates (Right Ascension: 05:34:39.10; Declination: +28:03:03.78). This TPF ensured light capture predominantly from the target star and its immediate surroundings.

Subsequently, the \texttt{Lightkurve} package was used to generate the TPF data's light curve. The extraction process involved defining an aperture mask based on a threshold of 3 standard deviations above the median flux level to select pixels corresponding to the target star's flux. This approach effectively isolated the star's light and reduced contamination. We estimated and subtracted the background flux from the target using masked pixels surrounding it. This procedure ensured that the resulting light curve was intrinsically related to V1371\,Tau. 

Figure\,\ref{fig:tpf} displays the application of the aperture mask and background subtraction, presenting the TPF alongside the selected aperture mask used for flux extraction. Additionally, black points indicating field sources from the \textit{Gaia} catalog with magnitudes $G < 13$ are overlaid to demonstrate that the target’s flux is free from contamination by nearby objects of similar brightness. Figure\,\ref{fig:V1371Tau_lc} shows the resulting light curve of V1371\,Tau as observed by TESS over the relevant sectors, with flux plotted as a function of BTJD (BJD$-$2,457,000), the time format adopted by TESS.

Additionally, we use the Kilodegree Extremely Little Telescope (KELT) data \citep{pepper2007kelt, pepper2012kelt}, a ground-based photometric survey optimized for detecting transiting exoplanets around bright stars ($8 < V < 10$), but also monitors stars in the $7 < V < 13$ range. It uses two 42\,mm wide-field telescopes located in Arizona (USA) and Sutherland (South Africa), with a broad $R$-band-like filter. KELT provides high-precision photometry ($\sim$7\,mmag for bright stars) over long baselines (up to 10 years) and a typical cadence of 30 minutes. Light curves are available in both raw and detrended versions, with detrending performed using the Trend Filtering Algorithm (TFA) as implemented in the VARTOOLS package. The data set is well-suited for studying Be star variability, probing the photosphere and inner disk regions.

\subsection{Spectroscopic Observations}

Our investigation boasts extensive spectral coverage of V1371\,Tau, mainly the visible ($\sim$0.40 - 0.90 $\mu$m) regions. All observation dates were converted to Barycentric TESS Julian Dates (BTJD = BJD - 2,457,000) to maintain consistency throughout the analysis. A summary of spectroscopic observations is presented in Table\,\ref{tab:specdata}.

High-resolution spectroscopic observations were obtained using the HERMES spectrograph \citep{raskin2011hermes}, a fibre-fed instrument on the 1.2m Mercator telescope at La Palma Observatory. This spectrograph offers a resolving power of R$\sim$85,000 and covers the wavelength range of $\sim$4000-9000\,{\text{\AA}}. We utilized 12 epochs of HERMES-reduced data collected between November 2020 and November 2023. Each epoch comprised a spectrum with an exposure time of 386 seconds. These observations were utilized to create the radial velocity curve of the eclipsing binary components of V1371\,Tau.

To classify the spectral types of the V1371\,Tau system (see Sect.\,\ref{sec:V1371Tauspec}), we utilized data from the Goodman Spectrograph \citep{10.1117/12.550069, crain2004goodman} mounted on the 4.1m SOAR telescope. Two observations were made with a spectral resolution ranging from R=1,400-14,000; the Goodman data were reduced using \textit{The Goodman Pipeline}, a Python-based software for producing wavelength-calibrated, science-ready, one-dimensional spectra. We used the Python tool \texttt{HANDY} for normalizing these spectra.

Finally, an analysis of the H$\alpha$ line variability was conducted using the BeSS database \citep{2011AJ....142..149N}, a rich repository maintained by the LESIA laboratory at the Observatoire de Paris-Meudon, which compiles thousands of spectra of Be and related stars from both professional and amateur contributors. For our study, we gathered medium- to high-resolution spectra of V1371\,Tau from 21 epochs between February 2009 and September 2024, using data from the BeSS database.

\begin{table}
   \footnotesize
   \centering
   \caption{Summary of spectroscopic observations of V1371\,Tau.}
   \label{tab:specdata}
   \begin{tabular}{cccc}
      \hline \hline
      \multicolumn{1}{c}{\multirow{2}{*}{Date}} & \multicolumn{1}{c}{\multirow{2}{*}{Source\,(qty.)}} & $t_{\rm exp}$ & R \\
       & & (sec.) & ($\lambda/\Delta \lambda$) \\
      \hline
      2020-11-18 & \multicolumn{1}{c}{\multirow{2}{*}{HERMES\,(12)}} & \multicolumn{1}{c}{\multirow{2}{*}{386}} & \multicolumn{1}{c}{\multirow{2}{*}{85,000}} \\
      to 2023-11-19 & & & \\
      2023-02-11 & \multicolumn{1}{c}{\multirow{2}{*}{Goodman/SOAR\,(2)}} & \multicolumn{1}{c}{\multirow{2}{*}{180}} & \multicolumn{1}{c}{\multirow{2}{*}{1,400–14,000}} \\
      and 2023-11-25 & & & \\
      %2022/12/06 & \multicolumn{1}{c}{\multirow{2}{*}{TripleSpec/SOAR\,(3)}} & \multicolumn{1}{c}{\multirow{2}{*}{240}} & \multicolumn{1}{c}{\multirow{2}{*}{3,500}} \\
      %to 2024/02/18 & & &  \\
      2009-02-21 & \multicolumn{1}{c}{\multirow{2}{*}{BeSS Database\,(22)}} & \multicolumn{1}{c}{\multirow{2}{*}{Var.}} & \multicolumn{1}{c}{\multirow{2}{*}{3,400-18,000}} \\
      to 2024-09-10 & & & \\
      \hline \hline
   \end{tabular}\\
   {Notes: Dates are given in YYYY-MM-DD format.}
\end{table}

%======================
\section{The Eclipsing Binary in V1371 Tau} \label{sec:binV1371Tau}
%======================

\subsection{Ephemeris and Eccentric Orbit Identification}

Using photometric data from TESS, we were able to accurately determine the ephemeris of the inner eclipsing binary system of V1371\,Tau. To eliminate any doubt, we consider that the primary component of the inner binary system is the star that is being eclipsed at the moment of greatest depth shown in the light curve, that is, at orbital phase zero. We employed two Gaussians to model the primary consecutive eclipses. We estimated their parameters using the Markov Chain Monte Carlo (MCMC) technique implemented in the Python package \texttt{emcee} \citep{foreman2013emcee}. Based on the mid-times of two consecutive primary eclipses available between TESS sectors 43 and 45, we derived a linear ephemeris with the initial epoch being the mid-time of the first eclipse. Subtracting the mid-time of the second eclipse yielded an orbital period of 33.62089 days. The linear ephemeris ($T$) is expressed as:
\begin{equation}
    T ({\rm BTJD}) = 2,490.00686 + 33.62089 \times E \, ,
    \label{eq:lin_efem}
\end{equation}
where $E$ is the cycle number from the initial epoch.

Upon phasing the light curve according to the derived ephemeris, we discovered that the secondary eclipse did not occur at phase 0.50 but closer to phase 0.35. This significant deviation from the expected phase in an EB system with a circular orbit confirms the presence of eccentricity. The observed secondary eclipse phase also suggests a periastron argument greater than $\pi/2$ radians. Performing a preliminary analysis of the light curve, we found a measurement of 0.268 for the eccentricity, a measurement that corroborates the result of 0.26 obtained by \citet{Rivinius2025}, and 150$^\circ$ for the periastron argument.

\subsection{Spectral Temperature and Rotational Broadening}\label{sec:V1371Tauspec}

Previous studies have classified V1371\,Tau with varying spectral types: B1Ve \citep{silk1973high, sallmen2004intermediate}, B1$\pm$0.5V  \citep{dinccel2015discovery}, and B0ne \citep{merrill1943supplement}. More recently, \citet{Rivinius2025} pointed out that the spectrum of V1371\,Tau appears to be that of a star of type B1IIIe. These results highlight the need for a more rigorous and consistent analysis, now taking into account mainly the fact that we have a combined spectrum of three stars. The best technique to perform such an analysis is spectral disentanglement. However, given the limited number and resolving power of spectra available, the most reliable approach is spectral reconstruction with theoretical models and comparison with the observed spectrum to obtain greater accuracy in determining the characteristics of the stars. This approach allowed us to determine the atmospheric parameters of all three components of the system, including effective temperature, surface gravity ($\log g$), and projected rotational velocity ($v \sin i$) \citep{hadrava1995orbital, shenar2020hidden}.

The presence of HeI $\lambda$\,4471$\text{\text{\AA}}$ closely aligned with MgII $\lambda$\,4481$\text{\AA}$ confirms the B-type nature of all the stars composing V1371\,Tau. Furthermore, its spectrum shows no characteristics of a cool B-type or early O-type star. We employ a synthetic spectrum that is the sum of three pre-computed synthetic energy distributions from TLUSTYBSTAR2006 \citep{lanz2007grid} with temperature, $\log\,g$ and $v \sin i$ as free parameters, and abundances, micro-turbulence and macro-turbulence as fixed parameters. The temperatures ranging from 18,000 to 30,000\,K (with 1,000\,K steps), $\log\,g$ between 3.50 and 4.75 (with steps of 0.25) and $v \sin i$ between 20 and 200\,km\,s$^{-1}$ (with 10\,km\,s$^{-1}$ steps), fitting it with the available spectrum from the $\chi^2$ method. We accessed the pre-computed models generated with \texttt{SYNSPEC} \citep{hubeny2011synspec} for spectral modeling. Solar abundances, a micro-turbulence velocity of 2\,km\,s$^{-1}$ and macro-turbulence of 25\,km\,s$^{-1}$ were assumed in all models \citep{simon2014iacob}. The error associated in measurements was calculated from the lowest $\chi^2$ within the 1$\sigma$ confidence interval. This is equivalent to approximately 3000\,K for temperature, 0.25\,dex for $\log\,g$ and 40\,km\,s$^{-1}$ for $v \sin i$.

Considering that the inner eclipsing system produces a greater variation in the wavelength shift, two Gaussians were previously adjusted, and their central wavelengths were set as the center of the theoretical models of the blueshift star and the redshift star. The outer component, in turn, because of a smaller disturbance in the wavelength shift, had its model centered at the laboratory wavelength. The Goodman spectrum, obtained at 2,986.5401\,BTJD and with an orbital phase of $\sim$0.77, was used in this analysis because of its with higher signal-to-noise ratio (S/N) spectrum ideal for fitting the line profiles. In this case, the primary component of the eclipsing system appears redshifted, while the secondary component appears blueshifted.

\begin{figure}[!ht]
    \footnotesize
    \centering
    % Answer: [trim={left bottom right top},clip]
    \includegraphics[trim={0.2cm 1.0cm 1.0cm 0.5cm},clip, width=\linewidth]{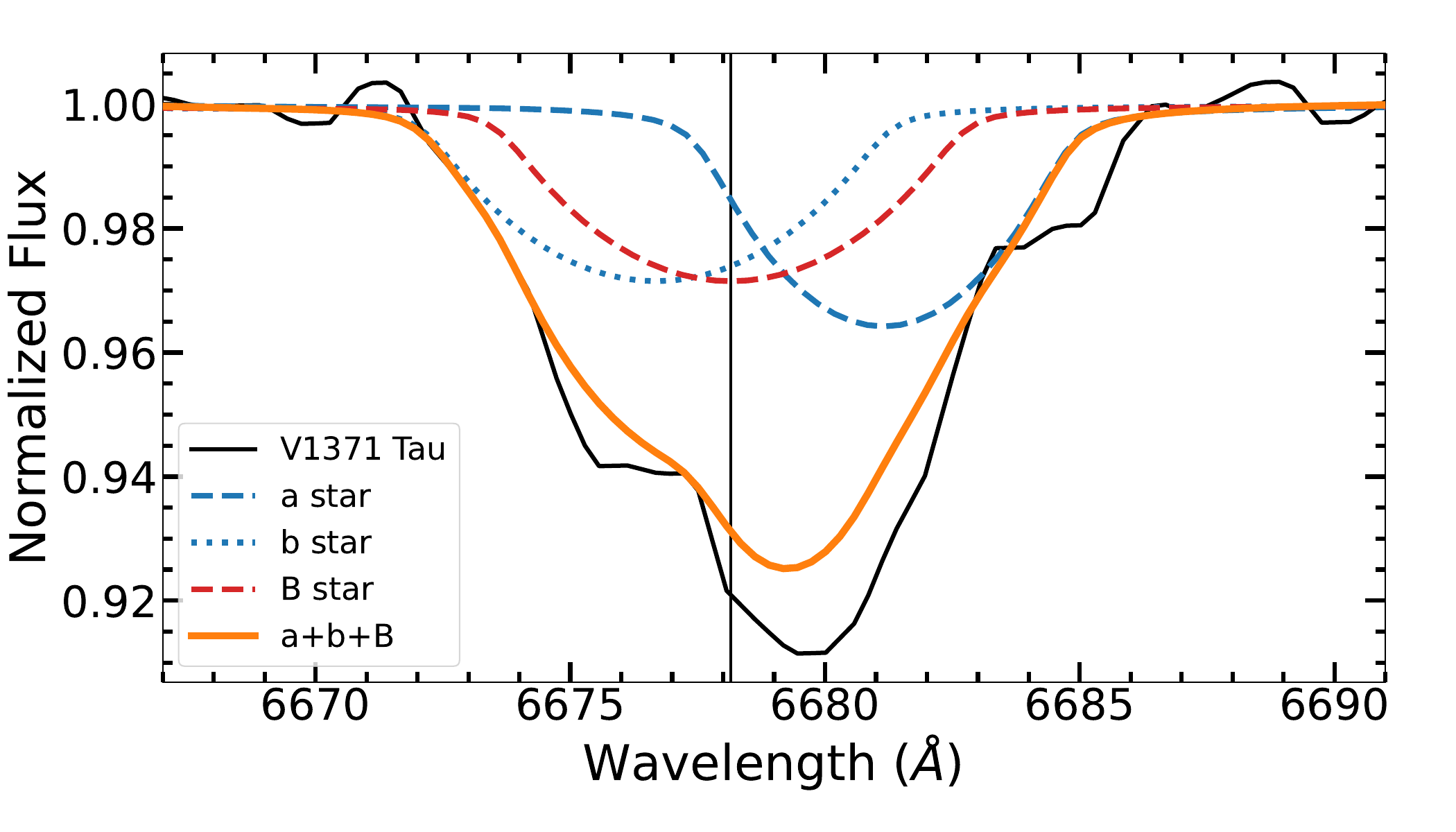}
    \caption{Comparison between the sum (orange solid line) of the best-fit TLUSTY spectral models for each component of V1371\,Tau  (non-solid lines), and the observed Goodman spectrum (black solid line) obtained on 2023-02-11. The laboratory wavelength of the HeI line, $\lambda$\,6678.152\,{\text{\AA}}, is indicated by the vertical black line.}
    \label{fig:spec3models}
\end{figure}

The model with the lowest $\chi^2$ and that reproduced the HeI line with the best accuracy is represented in Fig. \ref{fig:spec3models}, and resulted in the following values for each component of the system:
T$_{\rm a}$\,=\,25,000\,K, $\log\,g$$_{\rm a}$\,=\,4.50\,dex, $v_{\rm a}$\,$\sin$\,i\,=\,160\,km\,s$^{-1}$;
T$_{\rm b}$\,=\,28,000\,K, $\log\,g$$_{\rm b}$\,=\,4.50\,dex, $v_{\rm b}$\,$\sin$\,i\,$\geq$ \,200\,km\,s$^{-1}$;
T$_{\rm B}$\,=\,28,000\,K, $\log\,g$$_{\rm B}$\,=\,4.50\,dex, $v_{\rm B}$\,$\sin$\,i\,$\geq$\,200\,km\,s$^{-1}$.

\begin{figure}%[!ht]
    \footnotesize
    \centering
    % Answer: [trim={left bottom right top},clip]
    \includegraphics[trim={0.1cm 0.1cm 0.1cm 0.1cm},clip,width=\linewidth]{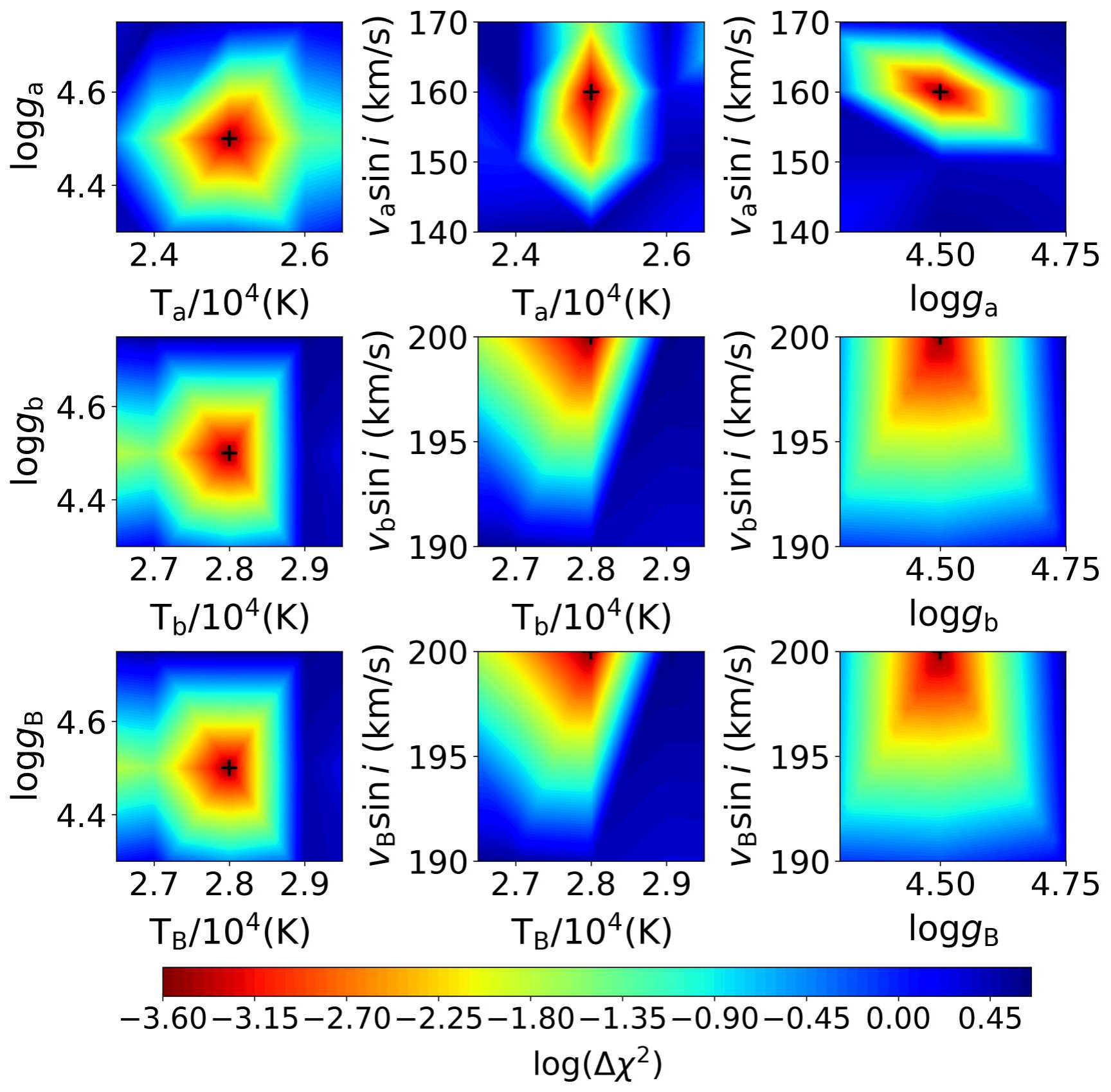}
    \caption{Distributions of effective temperature, surface gravity, projected rotational velocity, and $\chi^2$ values for the three stellar components, providing a view of both parameter discrete sampling strategy adopted in this analysis. The highlighted points mark the best-fit solutions (lowest $\chi^2$) for each parameter combination.}
    \label{fig:posspec3models}
\end{figure}

Figure\,\ref{fig:posspec3models} shows the distribution of $\chi^2$ values for the grid of synthetic composite spectra used to constrain the atmospheric parameters of the three stellar components. The upper limit of $v\sin i = 200$\,km\,s$^{-1}$ in the modeled grid reflects the limitations of the rotational broadening kernel implemented in \texttt{rotin-SYNSPEC}, which assumes rigid rotation and a linear limb-darkening law, and does not include gravity darkening, stellar oblateness, or line-profile distortions that become significant at higher rotation rates. These effects are particularly relevant for early-type rapid rotators, so extending the $v\sin i$ grid beyond this limit would not produce physically reliable line profiles within this formalism.

The best-fitting solutions correspond to spectral types B1 for component $a$, B0 for component $b$, and B0 for the external Be star $B$, based on the calibrations of \citet{bohm1981effective} and \citet{eker2020empirical}.

The composite spectrum of V1371\,Tau shows line wings broader than those reproduced by the synthetic multi-component model (Fig.\,\ref{fig:spec3models}), indicating the presence of a rapidly rotating contributor. To obtain an independent estimate of the rotational broadening, we measured the width of the He\,I $\lambda4471$\,\AA\ line following \citet{gray2005photospheres}, yielding $v\sin i = 250 \pm 40$\,km\,s$^{-1}$. For early-type B stars, reliable $v\sin i$ determinations commonly use strong, relatively unblended lines such as He\,I $\lambda4471$ or Mg\,II $\lambda4481$; in our case, He\,I $\lambda4471$ was preferred to minimize blending and wind contamination. We attribute this broadening to the external component $B$, consistent with its Be nature and the presence of a circumstellar disk. Other lines (e.g. HeI\,$\lambda$6678) were not used to ensure consistency and minimize blending and stellar wind effects. This independent approach is in line with the value of $v \sin i$ inferred through the spectroscopic fitting of the external Be component, and will be the value adopted as the one referring to it, given that this should be the largest contribution to the spectral broadening and that our model is limited to a Rotational Broadening of 200\,km\,s$^{-1}$.

Although the spectral lines of the inner components are blended in the composite spectrum, the fits do not require extreme line broadening or profile asymmetries indicative of near-critical rotation. Thus, while only a lower limit of $\sin$\,i\,$\geq$ \,200\,km\,s$^{-1}$ can be established for component $b$, the data are consistent with a rotation rate moderately higher than that of component $a$, but still well below the near-critical regime. This is expected for early-type B stars that are not Be stars.

Stellar luminosities are a key parameter for our subsequent analyses. Based on the derived atmospheric parameters, we estimate the luminosities using the following relation from \citep{gray2005photospheres}:
\begin{equation}
    \log \left( \frac{\rm{L}}{\rm{L}_\odot} \right) = 4 \log \left( \frac{\rm{T}_{\text{eff}}}{\rm{T}_\odot} \right) - \log \left( \frac{g}{g_\odot} \right) + \log \left( \frac{\rm{M}}{\rm{M}_\odot} \right) \, ,
\end{equation}
where, assuming M = 14\,M$_\odot$ for all stars, T$_\odot$ = 5,777\,K, and $\log g_\odot$ = 4.437\,(cgs), we have:
\begin{itemize}
  \item Star a: $\log (\rm{L}/\rm{L}_\odot) \approx 3.63$
  \item Stars b, B: $\log (\rm{L}/\rm{L}_\odot) \approx 3.82$
\end{itemize}

\subsection{Radial Velocity Scheme}\label{sec:rv}

The orbit of V1371\,Tau has never been previously constrained. In the spectra with emission from the H$\alpha$, we observe absorption lines that repeat in some lines associated with HeI. They present a cyclic variation with a period of $\approx$\,34 days, similar to the orbital period obtained in the light curve; therefore, such a variation is associated with the inner eclipsing binary system and is not associated with the Be star. We then find a new indication that V1371\,Tau is a triple system since the emission lines in H$\alpha$ and some HeI lines, referring to the Be star, do not exhibit orbital motion with the same period as the inner eclipsing binary. The observed peak separation ($\Delta\lambda \approx 9.5$\,\text{\AA}, corresponding to $\Delta v \approx 430$\,km\,s$^{-1}$) in the HeI emission lines is significantly larger than the $v \sin i$ of the inner binary components, suggesting that the emission originates in a circumstellar disk around the outer star rather than from the stellar surfaces of the inner binary.

We independently derive it here following a radial velocity (RV) measurement technique from Gaussian fitting and Doppler calculation, finding that it is a double-lined spectroscopic binary system (SB2). We followed a similar procedure presented in \citet{shenar2022x, rocha2022distance}. In short, gaussian profiles were fitted for the eclipsing components to the HERMES data for the HeI\,$\lambda$\,4921.931\,\text{\AA} and in particular for the HeI\,$\lambda$\,6678.152\,\text{\AA} absorption and emission lines, where the fits can be observed in detail in the figure\,\ref{fig:rv6678}. We incorporated the spectral analysis of the B component from Sec.~\ref{sec:V1371Tauspec} into a three-component fitting model, which even being a stationary or slowly varying component, can influence the shape and depth of the combined line profiles, affecting both the line intensity and potentially biasing the RV measurements. The wider profile of component B in Fig.\,\ref{fig:rv6678} arises due to this difference in the models, but since the radial velocities are obtained from the Gaussian centroids, the wider profile of B has no impact on the measured velocities or their uncertainties. As can be seen, the variation in the wavelength of the emission lines (which we associate with the outer component) is irrelevant. Its explained by its orbital period on a scale of years. Its presence does not generate systematic errors in the measurement of the radial velocity of the eclipsing binary system. We adjusted the absorption line profiles by fixing the depth and FWHM of each component throughout all epochs. To identify the lines of the primary and secondary components, we used the fact that we do not have spectra in orbital phases smaller than 0.13 or larger than 0.82, and the radial velocity lines should cross around orbital phase 0.35. Heliocentric correction was applied to the measurement scale of all RVs. The \texttt{emcee} code was used to search for the best solution of the models and sample the parameter space to obtain their respective uncertainties. The best value of the central wavelength of each Gaussian was used to calculate RV from the Doppler effect and its respective errors.

\begin{figure*}
    \centering
    % Answer: [trim={left bottom right top},clip]
    \includegraphics[trim={0 0.15cm 0.1cm 0},clip, width=6.2cm]{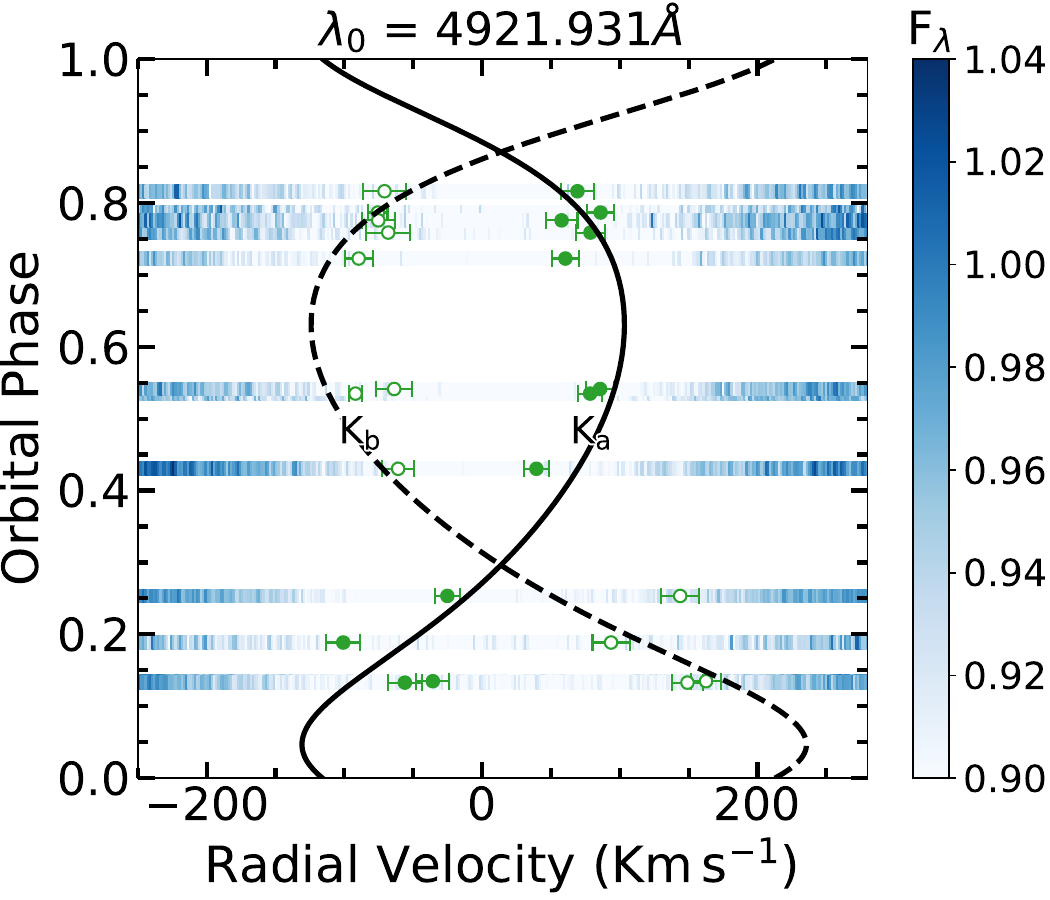}
    \includegraphics[trim={0 0.15cm 0 0},clip, width=5.4cm]{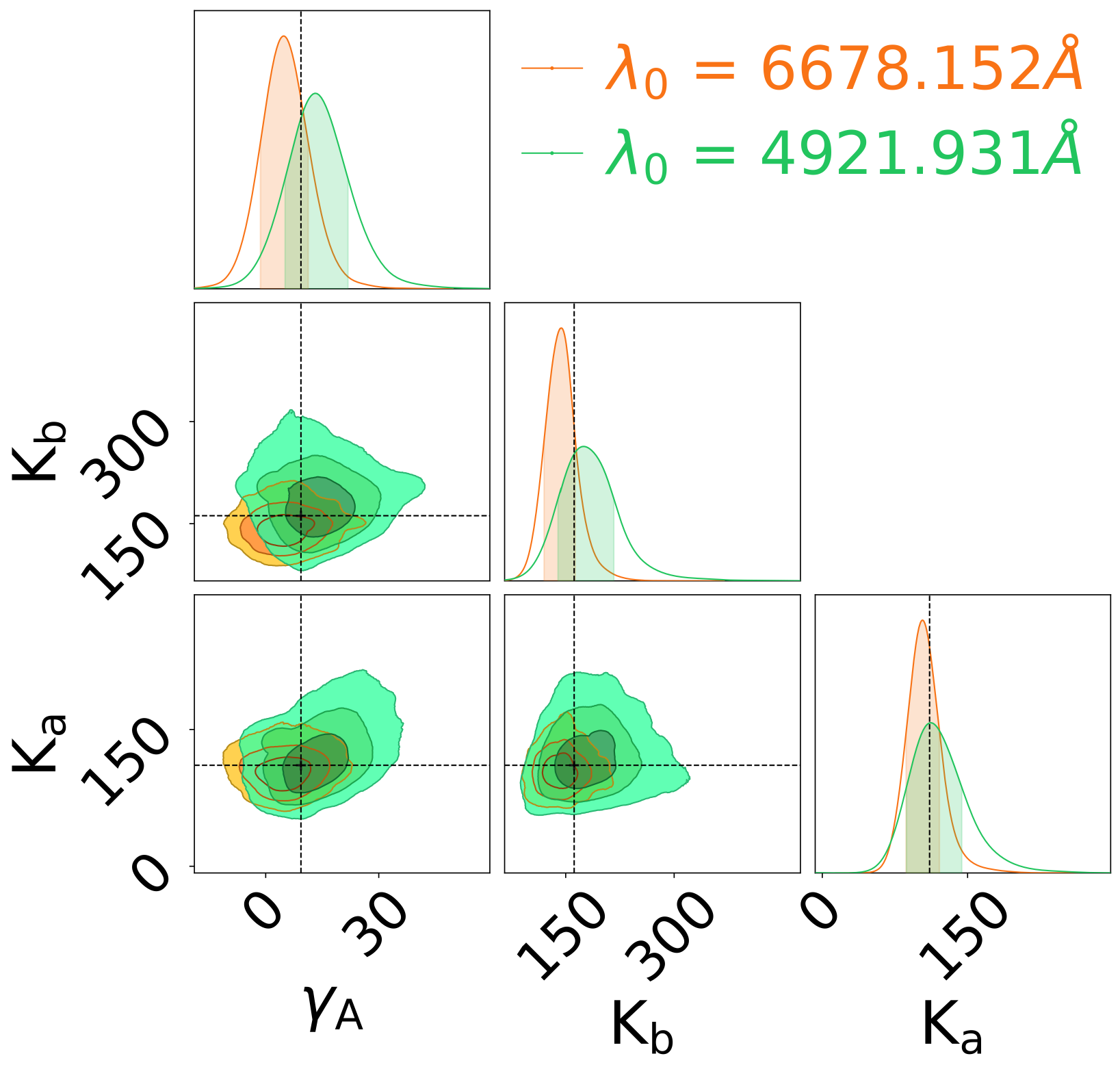}
    \includegraphics[trim={0 0.15cm 0.1cm 0},clip, width=6.2cm]{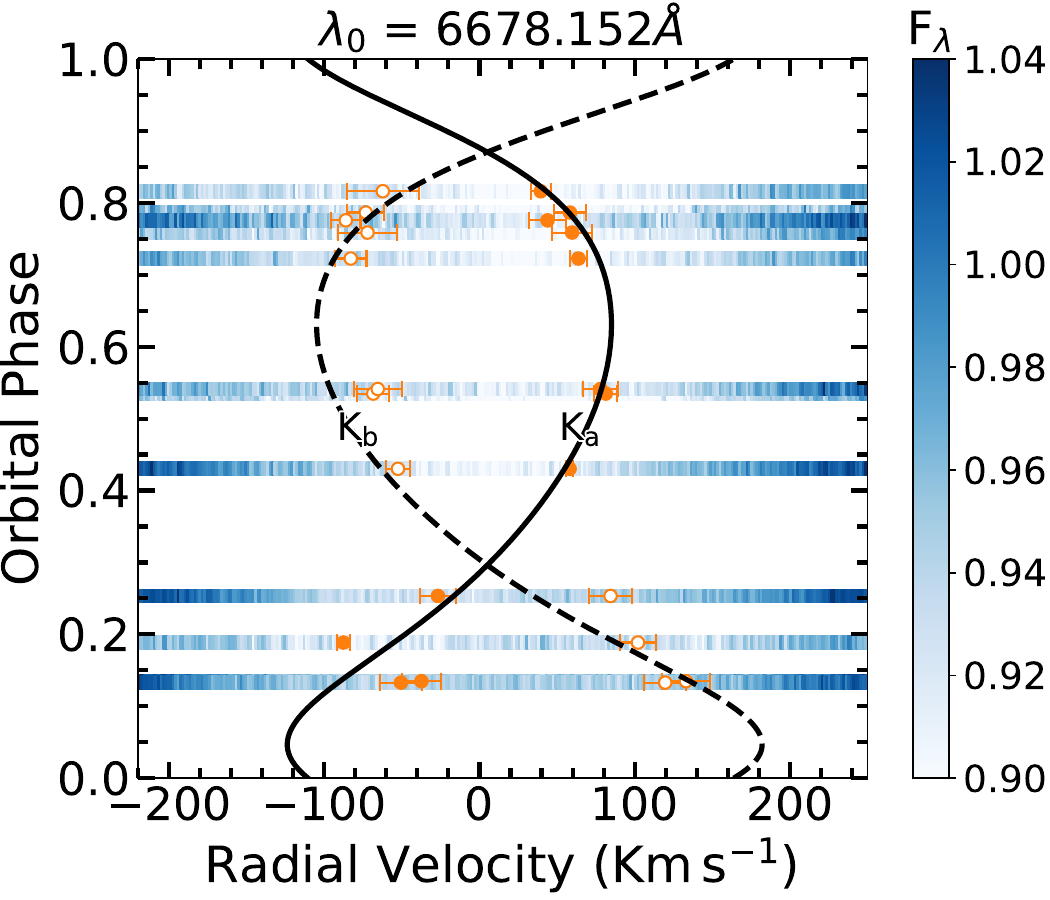}
    \\ \vspace{0.25cm}
    \caption{Left and Right Panel: The dynamical spectra of the HeI\,$\lambda$4921.931\,{\text{\AA}} and HeI\,$\lambda$6678.152\,{\text{\AA}} in V1371\,Tau, phased with the system’s orbital period of P\,=\,33.6209\,d. Colour-scales indicate the continuum-normalised flux F$_\lambda$. The orange and green dots are the observed points with their respective error bars, and the black solid line shows the radial-velocity curve model of the system.; Middle Panel: A corner plot generated using the \texttt{emcee} sampler illustrating one- and two-dimensional posterior distributions of free parameters derived from the Eq. \ref{eq:rv} model fitted to the V1371\,Tau inner binary radial velocity points for HeI\,$\lambda$4921.931\,{\text{\AA}} and HeI\,$\lambda$6678.152\,{\text{\AA}} lines.}
    \label{fig:rv}
\end{figure*}

Initially, we used the following synthetic model of RV's suitable for our observed points:
\begin{equation}
    \rm{RV}(\nu) = \gamma + K (\cos(\omega + \nu) + e\cos\omega ) \, ,
    \label{eq:rv}
\end{equation}
where $\nu$ is the true anomaly. We performed simulations for RVs measured using the Eq.\,\ref{eq:rv}, and tested it with the help of \texttt{emcee}, fixing the eccentricity and periastron argument with the previous light curve analysis and result. The best fit obtained is shown in Figure\,\ref{fig:rv}. To avoid including error in our following procedure, we decided to use the RV coming from the best fit, that is, the RV coming from HeI\,$\lambda$\,6678.152\,{\text{\AA}} line, with the RV gaussian fitting for each line available in Ap.\,\ref{ap:rv}. The average of the derived orbital elements (with 68\% confidence interval) agrees well with the expectations, the systemic velocity being: $\gamma_{\rm A}$\,=\,3.24$^{+5.75}_{-5.42}$ \,km\,s$^{-1}$ and finally the half-amplitudes of the velocity: K$_\mathrm{a}$\,=\,101$^{+16}_{-14}$\,km\,s$^{-1}$ and K$_\mathrm{b}$\,=\,140$^{+22}_{-20}$\,km\,s$^{-1}$.

\subsection{Binary modeling Scheme} \label{sec:bms}

We adopt a light curve modeling approach similar to those described \citet{rocha2022distance} to characterize the eclipsing binary V1371\,Tau. This procedure aims to derive crucial physical and orbital parameters through light curve modeling, encompassing parameters such as eccentricity, periastron argument, inclination, gravitational potentials for both components, temperature ratio, mass, and radii of both inner stars. The Wilson \& Devinney code (WDC), initially delineated by \citet{wilson1971realization} and subsequently adapted to accommodate the TESS photometric pass-band ($T$-band) in its 2019 version, is used to generate synthetic light curves.

To optimize the search for the light and RV curves that best align with our observational data, we utilize the Python package \texttt{emcee}, employing an MCMC ensemble sampler for a statistical comparison between synthetic and observational data.

In the TESS light curve, the eclipse depth for the primary star is $\approx$\,17\%, while for the secondary star it is $\approx$\,16\% of the normalized flux, including the constant contamination from the third component flux. As discussed in Section\,\ref{sec:V1371Tauspec}, the third star contributes roughly 38\% of the total system light, which leads to a dilution of the apparent eclipse depth. To find the true eclipse depth of the eclipsing stars alone, we remove the flux contribution of the third component from the light curve. The corrected eclipse depth ($D_{\text{corrected}}$) is given by:
\begin{equation}
    D_{\text{corrected}} = 1 - \frac{F_{\text{ecl}} - F_3}{F_{\text{total}} - F_3} \, ,
\end{equation}
where: $F_{total}$ = 1.00 (flux out of eclipse), $F_{ecl}$ = 0.83 (flux during primary eclipse), $F_3$ = 0.38 (flux from third component)

After applying this correction, the true primary eclipse depth increases from $\sim$17\% to $\sim$27\%, indicating that the actual flux variation caused by the eclipsing pair is substantially deeper than initially measured.

For both the WDC and the \texttt{emcee}, the input comprises the light curve derived from TESS data and RV from HERMES data, ensuring that its point sampling comprehensively covers the entire orbital phase of the system. The prior parameters, deduced in the preceding section, serve as initial inputs, except inclination and gravitational potentials. These parameters are obtained exclusively through the analysis of the light curve. Establishing the initial epoch and orbital period according to Eq.\,\ref{eq:lin_efem}, we derive their initial values to serve as priors for the subsequent \texttt{emcee} optimization process. Since the system has a non-zero eccentricity and periastron argument, the initial time computed by the WDC occurs at the point of minimum separation of the system components, which is slightly different from the time calculated using the primary eclipse of the observed data. Thus, a previous analysis using the WDC was performed, defining the value of -0.067 for an orbital phase shift of the light curve.

Usually, the approximate similarity in primary and secondary eclipse depths suggests comparable temperatures for both stars. However, this does not apply to all cases, considering systems with non-zero eccentricity and inclination different from 90$^\circ$, which is the case for V1371\,Tau. We fixed the effective temperature of the primary component and used the previous value of the secondary component (T$_{\rm b}$) determined in sec.\,\ref{sec:V1371Tauspec} as an initial value. Finally, as in the RV analysis, we fixed the eccentricity and the periastron argument to avoid large variations in the shape of the EB orbit. In the eclipsing binaries, a crucial parameter is the mass ratio (q$_{\rm A}$\,=\,M$_{\rm b}$/M$_{\rm a}$\,=\,K$_{\rm a}$/K$_{\rm b}$), typically derived from the radial velocity amplitudes, and in our case resulting in a mass ratio: q$_{\rm A}$ = 0.72. As this measurement was deferred to the previous section, we let it free in modeling the light and RV curve with WDC to obtain the other parameters with greater precision.

The spectrum does not present any characteristics different from a B star of early subtype and therefore, we expect that not only the stars of the inner pair, but the third component also has parameters corresponding to a hot B-type star, so due to the B-type component's spectral type, both stars are assumed to possess radiative atmospheres. Consequently, gravity-dimming exponents ($g_{\rm a}$\,=\,$g_{\rm b}$) and albedos (A$_{\rm a}$\,=\,A$_{\rm b}$) were fixed to standard hot star values of 1.00 and 0.50, respectively \citep{claret2011gravity}. We adopted the square root limb darkening law in the WDC as recommended by \citet{van1993new}. We used the normalized light curve of the sectors observed in 2021 to isolate the system parameters to determine the sporadic phenomenon of brightness decay observed in the most recent sectors. Based on the observed morphology of the light curve, characterized by two plateaus and well-defined deep eclipses, we configured the Wilson-Devinney code in mode 2 (detached) to avoid restrictions on Roche lobe filling. Subsequently, we used \texttt{emcee} to generate a sample of 100,000 models and infer the best values and associated error bars for the following parameters:

\begin{itemize}
    \item [] (i) Separation (a$_{\rm A}$);
    \item [] (ii) mass ratio (q$_{\rm ba}$).
    \item [] (iii) orbital inclination (i$_{\rm A}$);
    \item [] (iv) gravitational potentials of the primary ($\Omega_{\rm a}$) and secondary components ($\Omega_{\rm b}$);
    \item [] (v) effective temperature of the secondary component (T$_{\rm b}$).
\end{itemize}

In special, we have two independent measurements for T$_{\rm b}$. Using Bayes' Theorem and assuming that each measurement has a Gaussian distribution, we perform the Bayesian fusion of normal distributions in order to obtain a single measurement for T$_{\rm b}$, resulting in the combined value of 28,000\,K with an uncertainty of $\pm$2,300\,K.

\begin{figure*}
    \centering
    \includegraphics[width=\linewidth]{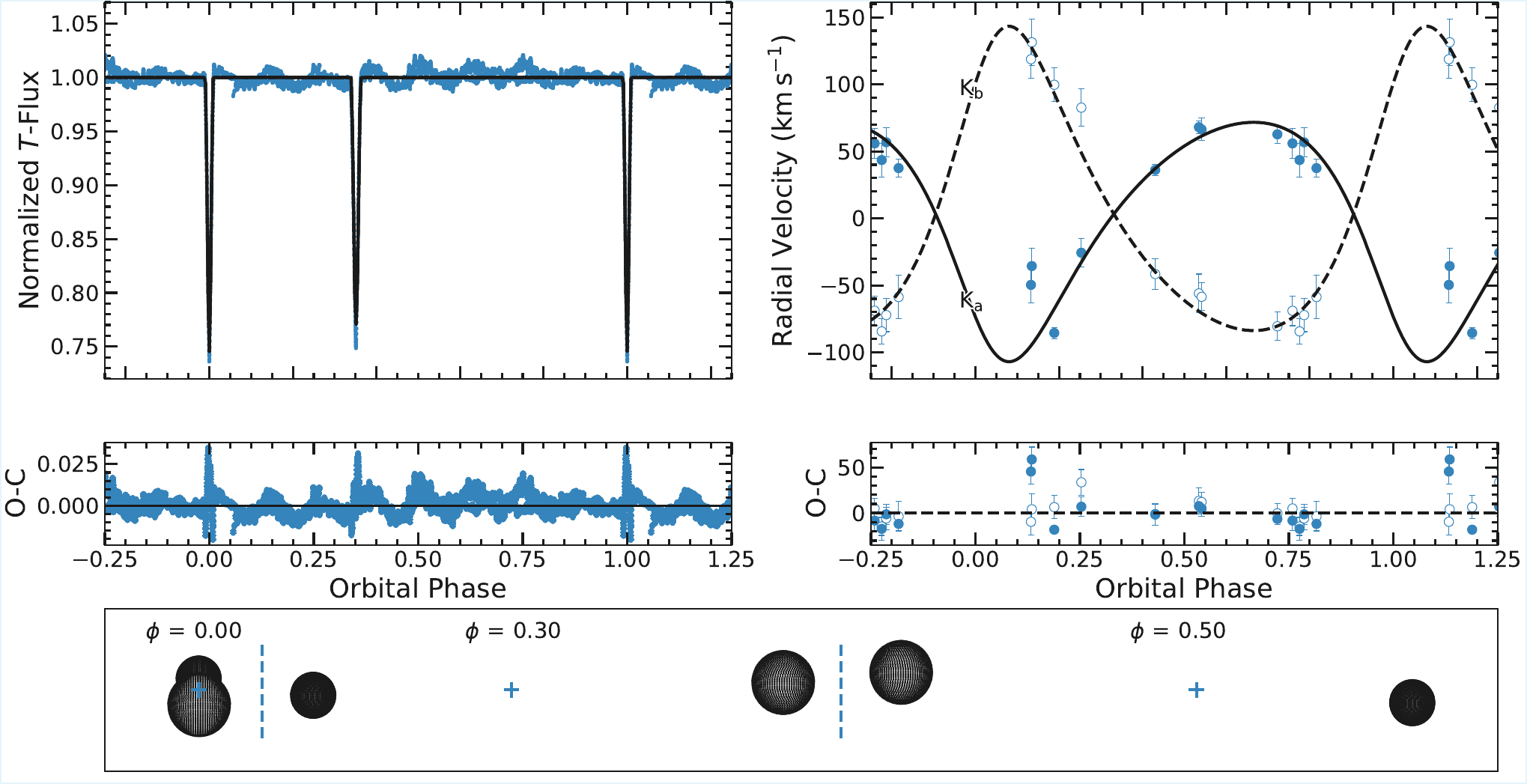} % Answer: [trim={left bottom right top},clip]
    \caption{Top panel: Light curve and radial velocity of eclipsing binary in orbital phase of The V1371\,Tau. The light contribution of the third component was removed from the normalized $T$-flux. The blue dots are the observed data, and the solid black line is the best-fit model generated by WDC.; Bottom panel: Unscaled 3D models of the eclipsing system in V1371\,Tau in three distinct orbital phases, showing the relative positions of the stars and the system's center of mass (marked by a blue cross).}
    \label{fig:V1371Tau_ecl}
\end{figure*}

\begin{figure}
    \centering
    \includegraphics[width=\linewidth, trim={0.2cm 0.2cm 0.2cm 0.2cm},clip]{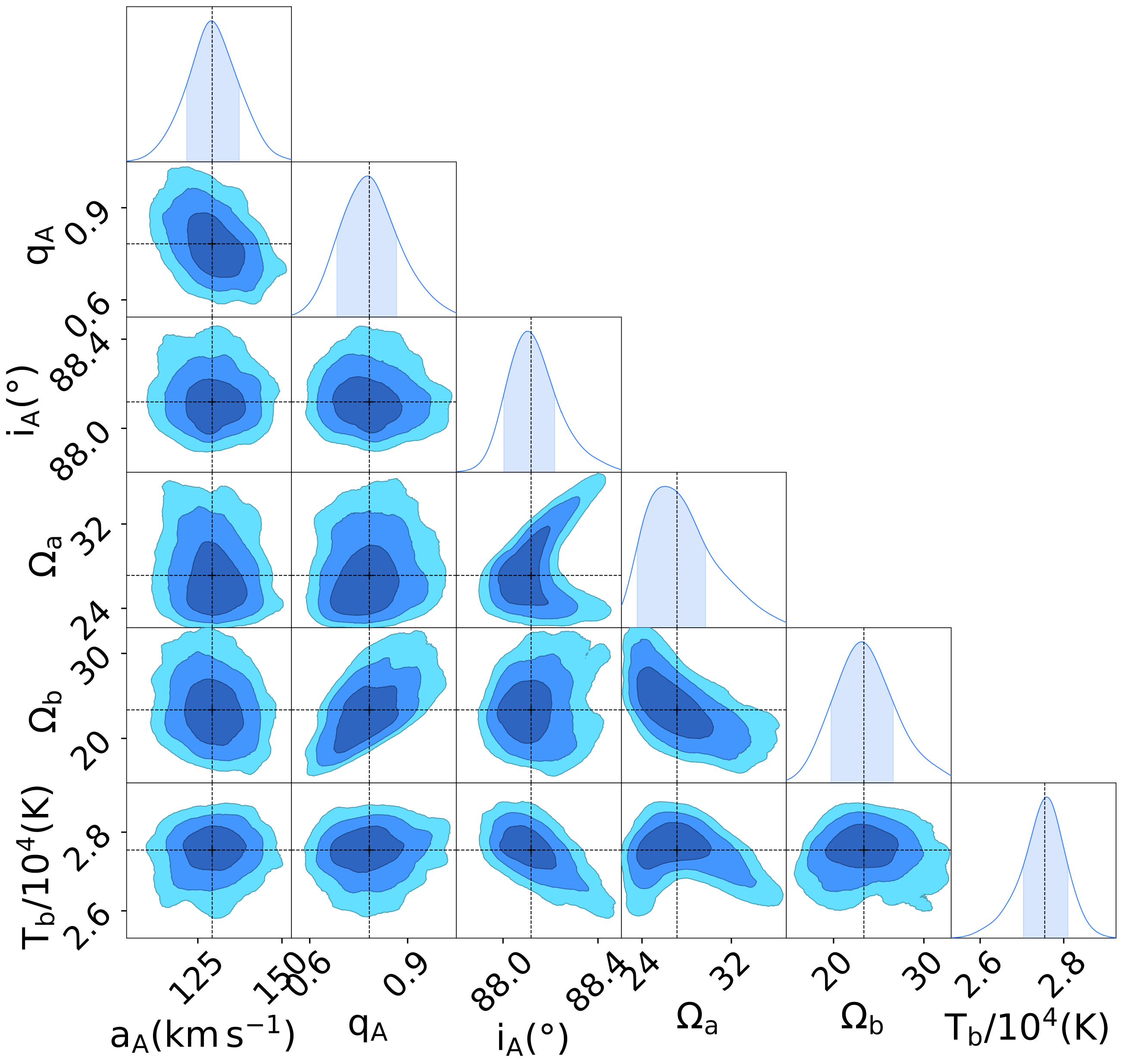} % Answer: [trim={left bottom right top},clip]
    \caption{Corner plot generated using the \texttt{emcee} sampler, showing the one- and two-dimensional posterior distributions of 100,000 iterations for the free parameters derived from the WDC models fitted to the V1371\,Tau. The plot represents the probability density of each parameter, with denser contours indicating regions of greater probability within the overall posterior distribution. The shaded colour regions correspond to 68\%, 95\%, and 99\% confidence intervals.}
    \label{fig:posteriori}
\end{figure}

Figure\,\ref{fig:V1371Tau_ecl} showcases the best-fitting model for the light curve and radial velocity exclusively of the eclipsing pair, below a scaled diagram that illustrates the relative sizes of the stars and the orbital motion, and the Fig.\,\ref{fig:posteriori} illustrates one-dimensional and two-dimensional posterior distributions of free parameters generated using \texttt{emcee}. Table\,\ref{tab:parameters} summarizes the orbital and physical final values and their uncertainties, obtained by averaging and considering the 1-$\sigma$ range directly adjusted. We note that the $\log g$ values inferred from the spectral fitting (Section~\ref{sec:V1371Tauspec}) differ slightly from those obtained through the binary modeling (Section~\ref{sec:bms}) what is expected, since the $\chi^{2}$ maps are based solely on line-profile fitting and are affected by degeneracies between $T_{\rm eff}$, $\log g$, and $v\sin i$, which are enhanced in composite spectra where the contributions from the individual stars are blended. In contrast, the $\log g$ values from the eclipsing binary solution are derived from dynamically constrained masses and radii, and are therefore more robust. For this reason, we adopt the $\log g$ values from Sect.\,3.4 as the final ones, noting that both estimates agree within the 1$\sigma$ uncertainties.

\begin{table}
\begin{center}
\caption{Resume of parameters for the eclipsing binary in V1371\,Tau.}
\label{tab:parameters}
% table
    \begin{tabular}{l c c}
    \hline
    \hline
    \multicolumn{1}{l}{\multirow{2}{*}{Parameters}} & \multicolumn{2}{c}{Inner Pair} \\
    & a & b \\ \hline
    Orbit Period (d) & \multicolumn{2}{c}{33.62090$^\dagger$} \\
    Time prim. ecl. (BTJD) & \multicolumn{2}{c}{2,490.00686$^\dagger$} \\
    Eccentricity & \multicolumn{2}{c}{0.268$^\dagger$} \\
    Long. of periastron ($^{\circ}$) & \multicolumn{2}{c}{150$^\dagger$} \\
    Syst. velocity (km\,s$^{-1}$) & \multicolumn{2}{c}{3.24$^{+5.75}_{-5.42}$} \\
    RV amplitude (km\,s$^{-1}$) & 101$^{+16}_{-14}$ & 140$^{+22}_{-20}$ \\
    Mass Ratio & \multicolumn{2}{c}{0.78$\pm$0.08} \\
    Inclination ($^{\circ}$) & \multicolumn{2}{c}{88.11$^{+0.12}_{-0.08}$} \\
    Surface potential & 27.13$^{+3.66}_{-2.55}$ & 23.36$^{+3.44}_{-2.05}$ \\
    Orbital ph. shift & \multicolumn{2}{c}{-0.067$^\dagger$} \\
    Eff. temperature (K) & 25,000$\pm$3,000$^\dagger$ & 28,000$\pm$2,300 \\
    Semi-major axis (R$_{\odot}$) & \multicolumn{2}{c}{129$^{+8}_{-7}$} \\
    Stellar mass (M$_{\odot}$) & 14$\pm$3 & 11$\pm$2 \\
    Stellar radii (R$_{\odot}$) &  4.9$\pm$0.3 & 4.8$\pm$0.6 \\
    Surface gravity (dex) & 4.21$\pm$0.11 & 4.16$\pm$0.11 \\
    Luminosity (log L/L$_\odot$) & 3.93$\pm 0.12$ & 4.11$\pm$0.11 \\
    \hline
    \hline
    \end{tabular} \\
    Notes: $^\dagger$Parameters fixed in all procedures and analyses.
\end{center}
\end{table}

\section{The Third Body and its Be Nature} \label{sec:beV1371Tau}

Previous distance estimates for V1371\,Tau ranged from 0.87\,kpc \citep{mendoza1958spectroscopic} to 0.90\,kpc \citep{gvaramadze2006supernova}. Leveraging the high-precision astrometry data from Gaia DR3, we have a significantly more precise distance of 1.20$\pm$0.03\,kpc with high reliability due to the Renormalized Unit Weight Error (RUWE) value\,$\leq$\,1.4. This result aligns well with the recent distance estimation ($\sim$1.30$\pm$0.1\,Kpc) for the nearby runaway star HD\,37424 by \citet{dinccel2015discovery}. This refined distance estimate forms the crucial basis for calculating the absolute parameters of the Be component of V1371\,Tau, such as separation, orbital period, and mass.

\begin{figure}
    \centering
    % Answer: [trim={left bottom right top},clip]
    \includegraphics[trim={5.0cm 3.75cm 1.20cm 5.2cm},clip, width=\linewidth]{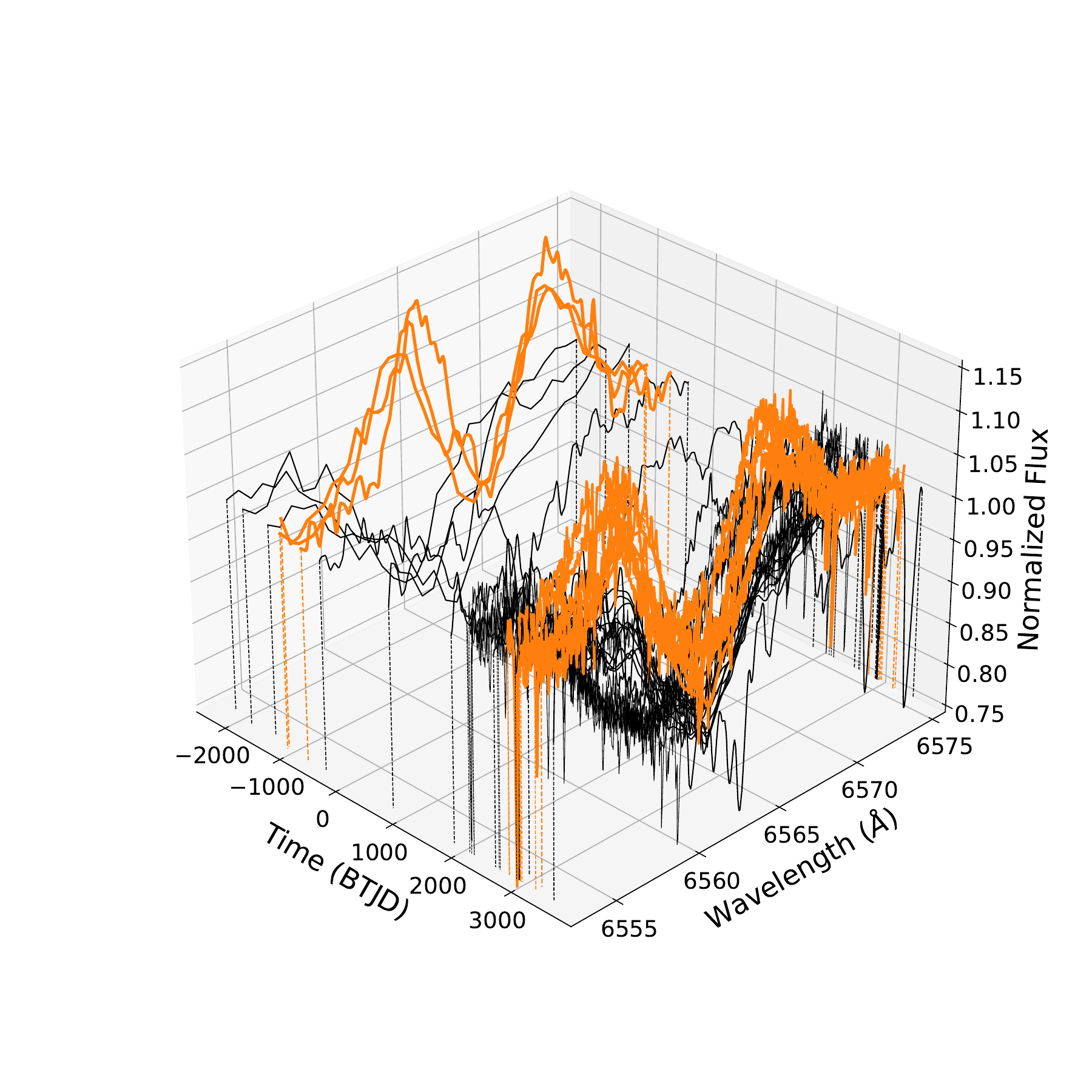}
    \includegraphics[width=\linewidth]{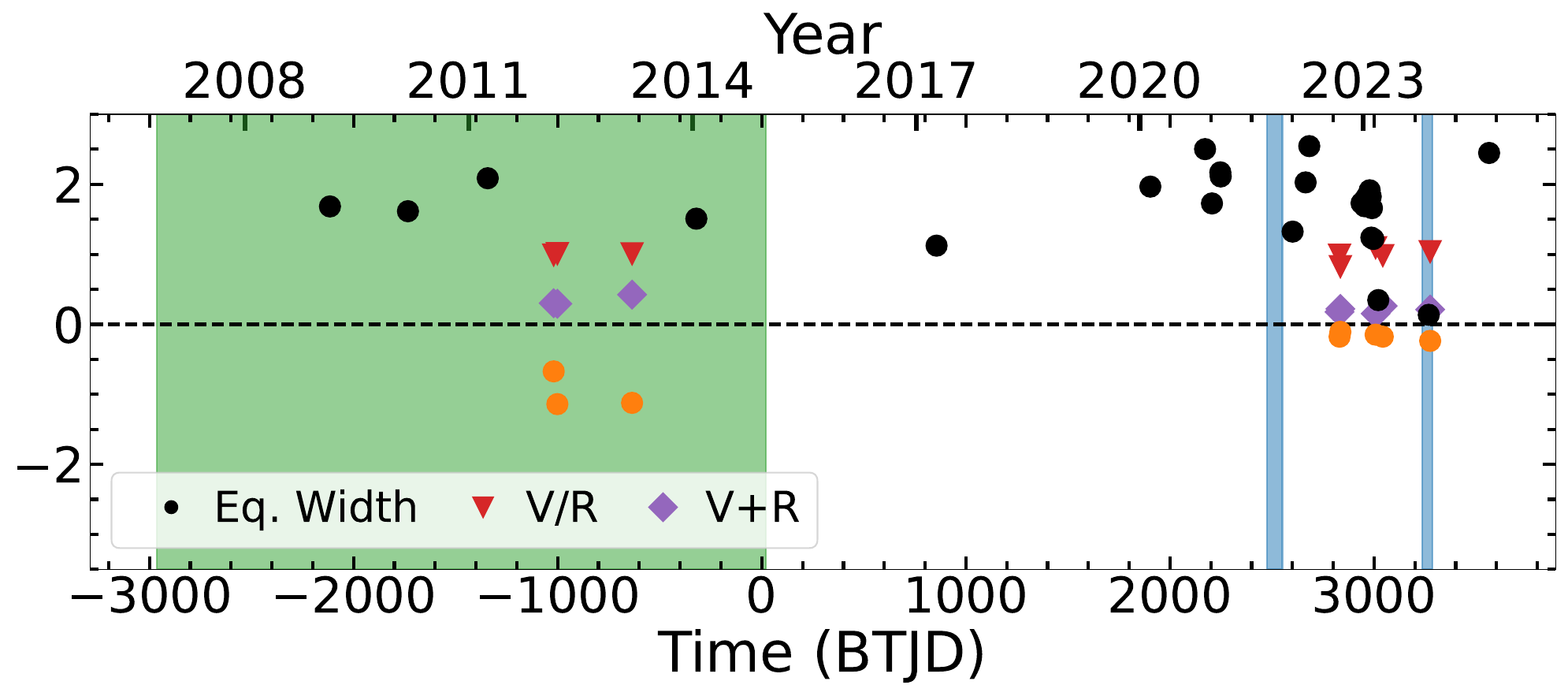} \\
    \vspace{0.15cm}
    \includegraphics[width=\linewidth]{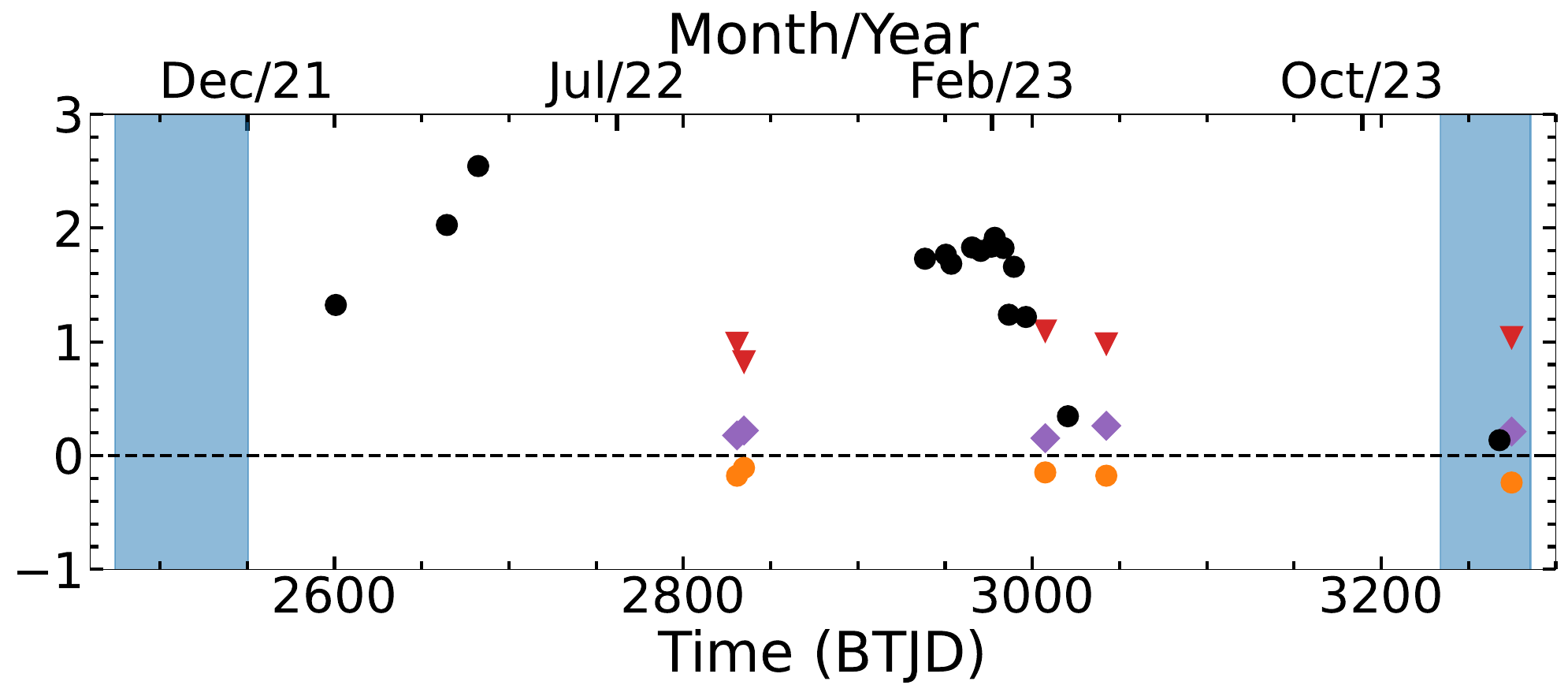}
    \caption{H$\alpha$ analysis of data from BeSS, Goodman and HERMES. Top Panel: Normalized H$\alpha$ emission lines from spectroscopic observations of V1371\,Tau between 2009 and 2024; Mid and Bottom Panel: EW (dot), V/R (inverted triangle) and V+R (Rhombus) of H$\alpha$. Orange dots indicate the EW of observations with H$\alpha$ emission. The blue band indicates the period of TESS photometric observations, while green represents KELT data. The bottom panel is a detail from the middle one.}
    \label{fig:ha}
\end{figure}

In Sec.\,\ref{sec:rv} we have already shown that H$\alpha$ emissions do not present a variation equivalent to a period of $\approx$34\,days, a new factor that corroborates with the interferometric measurement of 5\,$mas$ \citep{Rivinius2025}. Take the average of the dates with equivalent width (EW) of H$\alpha$ less than zero in 2012 and 2013, and subtract it from the average of the negative EW dates in 2023, we get a period of $\sim$11\,years. However, looking at the higher cadence H$\alpha$ data (in 2022/2023), we see variations on the time scale of months. Outside this time range, the data are too sparse to draw firm conclusions about the outer star orbital period. Furthermore, the V/R ratio is approximately always constant and around one (see Figure\,\ref{fig:ha}), so there is no physical mechanism and no evidence that the disk is perturbed by, for example, tidal forces from some other star.

V1371\,Tau appears to have a similar orbital configuration to the triple system $\nu$\,Gem, with the advantage that the inner binary system is eclipsing. Furthermore, the stars of V1371\,Tau are approximately four times more massive than the stars that make up $\nu$\,Gem. V1371\,Tau was initially categorized as a Be star by \citet{merrill1943supplement} due to its conspicuous hydrogen lines. To gain a comprehensive understanding of its Be nature, we conducted an in-depth examination of its historical H$\alpha$ emission patterns using archival data. Employing data from the BeSS catalog, the Goodman and HERMES spectrograph spanning the years 2009 to 2024, we discerned noteworthy H$\alpha$ emissions in 2012, 2013, and most recently, at 2023 (depicted in Fig.\,\ref{fig:ha}). Although KELT observed the star during the 2012–2013 H$\alpha$ events, any photometric changes associated with disk growth or outbursts were detectable. This intermittent H$\alpha$ emission pattern conforms to the distinctive mass ejection events characteristic of Be stars. Analysing the orbital phases of the inner binary with the observed H$\alpha$ emissions, we found that these emissions do not occur due to a transfer of matter caused by the approach of their eclipsing binary companions, which occurs around the inferior conjunction phase of $\sim$0.35. This indicates the true classical Be nature of the system, since this emission is caused by the formation of a Keplerian disk by the third body and that, considering the dilution of the flux from the triple system, the H$\alpha$ behavior of the Be component in this system appears typical of an early-type Be star and, it behaves like a single Be star due to its large separation from the center of mass of the inner system, $\sim$6\,AU, which is greater than 0.5\,AU, as proposed by \citet{abt1984stars}.

Although simultaneous TESS and H$\alpha$ data are not available for 2021, the consistently positive H$\alpha$ equivalent widths throughout the TESS observing period suggest that there was no major circumstellar disc present, nor significant mass accretion or decretion activity at that time. However, the TESS light curve shows long-term variations, which may be related to disc growth (during sectors 43/44/45) and dissipation (during sectors 71/72). In addition to the eclipses of the inner binary, the light curve also exhibits clear short-term variability outside eclipses. These signals are likely due to stellar pulsations, which may originate from one or more components of the system. Given that all Be stars are known to be pulsators, it is reasonable to attribute at least part of this variability to NRPs from the Be star.

Fortunately, we have extensive spectral coverage from 2023 and recent TESS observations from sectors 71 and 72. The analysis of the H$\alpha$ equivalent width reveals a period of intense variability and circumstellar activity in early 2023, likely associated with the erratic formation or growth of the Be star's disk. This may have included outbursts that caused an increase in the system's flux, followed by a decay observed at the beginning of sector 71, possibly due to the gradual dissipation of the disk. This scenario is supported by the 2024 spectral observations, which no longer show H$\alpha$ emission.
We cannot rule out the possibility of accretion of material from the SNR by the stars in the system. However, it is unlikely that this would cause a significant variation in the system, especially in the size of the Be star, since the system is on the edge of SNR S147, a place where the dust and gas from the original supernova have practically disappeared into the interstellar medium.

The coplanarity of the Be star with the inner eclipsing binary remains unrestricted. The absence of shell lines may suggest that the Be disk is not seen edge-on, although this is not conclusive, as weak shell features can be diluted by the contributions of the other components, and strong shell lines form only in stable, extended disks \citep{rivinius2006bright}. Furthermore, we do not have photometric coverage on the order of years to verify possible signs of eclipses caused by the Be component. Nevertheless, some quantitative conclusions can be drawn. The measured $v \sin i$ of the Be star is approximately 250\,km\,s$^{-1}$ (see Sec.\,\ref{sec:V1371Tauspec}). The inclination of the Be star remains unconstrained. Very low inclinations ($\lesssim 20^\circ$) appear unlikely given the broad line profiles observed, while very high inclinations ($\gtrsim 75^\circ$) would typically produce strong shell features that are not present in our spectra. The actual inclination, therefore, likely lies somewhere between these extremes, although this estimate is merely indicative and not conclusive. This discussion assumes that the stellar rotation axis is aligned with the orbital plane of the Be component, an assumption that cannot be directly tested with the available data. Adopting the measured $v \sin i$, we estimate the corresponding equatorial rotation velocity of $v_{\rm rot}\,\approx$\,290–500\,km\,s$^{-1}$ across this inclination range. From spectral modeling, and adopting typical calibrations for B0-type stars (M\,$\approx$\,14$\pm$3\,M$_\odot$ and R\,$\approx$\,6.2$\pm$1.5\,R$_\odot$), we estimate a critical velocity of $v_{\rm crit}\,\approx$\,656$\pm$80\,km\,s$^{-1}$. Although this is not based on direct measurements, it provides a reasonable reference for the Be star's rotation. Using the observed projected rotational velocity, we obtain an approximate critical rotation fraction:
\begin{equation}
    \frac{v_{\rm rot}}{v_{\rm crit}} \simeq 0.44\text{–}0.76 \,,
\end{equation}
where the range accounts for uncertainties in $v_{\rm crit}$ as well as possible inclination effects. Within these uncertainties, the Be star's rotation is consistent with the typical values observed for stars of its type.

The Be star's $v \sin i$ is significantly higher than that of the components of the inner eclipsing binary, which likely have inclinations close to $90^\circ$ and thus $v \sin i \approx v_{\rm rot}$. Therefore, we can confidently state that the Be star is rotating much faster than either component of the eclipsing pair. Moreover, the Be star's projected rotational velocity is consistent with typical values found in large Be star samples, especially if we consider the average value of $v/v_{\rm crit}$=0.65 \citep[e.g.,][]{zorec2016critical}.

This wide range reflects the strong dependence on the unknown inclination angle and projected rotational velocity. Although uncertain, this estimate indicates that the Be star is likely rotating at a substantial fraction of its critical velocity, consistent with expectations for Be stars. A more precise determination of $v/v_{\rm crit}$ will require spectral disentangling and modeling of the circumstellar disk to constrain both the inclination and the intrinsic stellar parameters more reliably.

\section{The Triple System V1371 Tau} \label{sec:V1371Tau}

The combination of the object's physical and orbital characteristics determined so far, together with the spectral classification of its brightest stars, reveals an intriguing puzzle in determining the configuration of this system. \citet{Rivinius2025} suggests the presence of a third component in the system with a separation of 5\,$mas$ (measurement obtained with GRAVITY Interferometry). In our analysis, we found other evidences that reinforces that this system is composed of three stars, the outer one being the classical Be star, like: i) the fact that the H$\alpha$ emission lines do not present a variation in wavelength consistent with the orbital period of the inner binary system, ii) show no sign of shell absorption, which would be expected from a Be star viewed at very high inclination angles (which is the case for the eclipsing pair), iii) The helium emission appears at velocities well outside the photospheric absorption lines of the two eclipsing stars, and iv) They show no asymmetric variation, which would be expected from a close eccentric binary, if one of them were a Be star.

Having a third body in a gravitational stellar system, the dynamical stability will depend on new parameters - no less than eleven - that are necessary to define its orbit and stability, such as those already known in single and binary stars (e.g., eccentricity), mainly separation, inclination, and the argument of the periastron of the stars in the inner and outer orbits. We will now briefly analyze the evolution based on the data and parameters obtained so far, using theoretical models of evolution.

\subsection{Evolutionary Status}

The separation and eccentricity of the inner binary system do not suggest continuous mass transfer. As discussed in the previous section, the presence of H$\alpha$ emission lines, even outside periastron (Fig.\,\ref{fig:ha}) indicates the Be nature of the system's third component.

\begin{figure}
    \centering
    % Answer: [trim={left bottom right top},clip]
    \includegraphics[width=\linewidth]{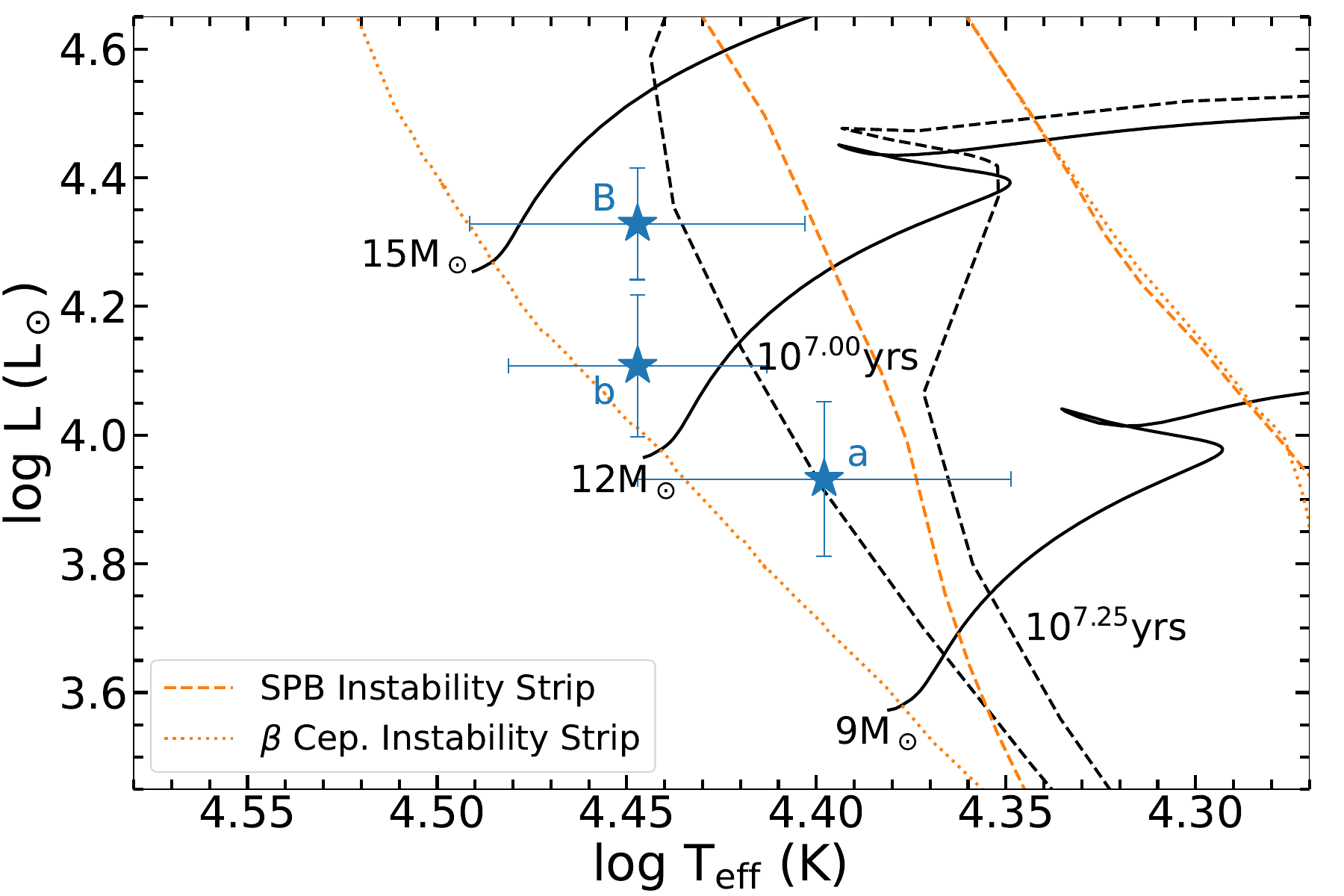}
    \caption{Hertzsprung-Russell diagram of V1371\,Tau based on solar-metallicity models from \citet{ekstrom2012grids} with $v$/$v_{\rm crit}$\,=\,0.40. Solid black lines indicate evolutionary tracks for different stellar masses, while dotted black lines represent isochrones of various ages. The system’s components are marked as blue stars.}
    \label{fig:hrdiagram}
\end{figure}

The Fig.\,\ref{fig:hrdiagram} shows the Hertzsprung–Russell Diagram (HRD) that was constructed to evaluate the evolutionary state of V1371\,Tau. The evolutionary and isochronous tracks are of \citet{ekstrom2012grids} for solar abundances and with rotation ($v$/$v_{\rm crit}$\,=\,0.40), and were compared with the parameters derived from the component stars of the triple system, represented in the figure by blue stars. To complement our study, we also present the instability domains of SPB and $\beta$\,Cepheid stars defined by \citet{2007CoAst.151...48M}.

The spectrum of V1371\,Tau gives us no indication of the presence of a star with a luminosity classification other than the main sequence, whether it is rich in hydrogen or helium. We can observe that the masses of the eclipsing components are not close to the theoretical trajectories, except the outer component Be. Considering the mass and radius measurements of the secondary component and its positions in the HRD, taking into account the error margins, we consider it as overluminous, even on the main sequence, similar to other objects already found in the literature \citep{rocha2022distance}. This characteristic may be linked to its rotation, since component b rotates faster than a, as if it had undergone a process of mass loss, altering its evolutionary stage, consequently making it more luminous than normal \citep{herrero92, deMink2009D}. Another fact is that this system is immersed in the supernova remnant S147, making the luminosity classification of this object a challenge given its possible influence on the brightness of the components due to reddening, which should be investigated further. Furthermore, the evolutionary models are based on single-star evolution, which introduces limitations when applied to systems with multiple stellar components. This result is also known in the literature due to the persistent discrepancy between the evolutionary and dynamical masses.

Analyzing the evolutionary status of the b star without assuming it is still in the core hydrogen-burning phase, one possible scenario is that it could be a post-EHB (Extended Horizontal Branch) object, eventually evolving into a “luminous” sdOB star. However, this interpretation is unlikely given the measured mass and radius. Furthermore, although the star lies within the $\beta$\,Cephei instability strip, it does not exhibit any characteristics of a $\beta$\,Cephei variable. In any case, the key to unveiling the evolutionary status of this system lies in spectral disentangling, which enables more robust and independent determinations of the atmospheric parameters of the individual components of V1371\,Tau.

The isochronous features that pass close to the main sequence components, we infer a typical age of $\sim$10\,Myrs for the stars of V1371\,Tau.

The system’s formation history raises two possibilities:
\begin{enumerate}
    \item The Be star may have evolved as an isolated rapid rotator, without past binary interactions.
    \item Alternatively, it may have undergone mass transfer from a now-undetectable companion (e.g., a faint sdOB star), which would imply a previous binary phase.
\end{enumerate}

However, current system mass estimates are insufficient to favor the second scenario, and no short-period luminous Be star companion was detected. Further spectroscopic and interferometric observations are essential to determine the Be star's mass, inclination, and potential unseen companions.

\subsection{Orbital Dynamics}

This system, initially studied with KELT ground-based photometry in 2011, revealed much more complexity when revisited with high-precision TESS data. Although KELT is capable of identifying primary eclipses with a depth of $\sim$0.17\,mag, none were observed \citep{labadie2017photometric}, probably because changes in the orbital geometry, as discussed below.

\begin{figure}
    \centering
    % Answer: [trim={left bottom right top},clip]
    \includegraphics[trim={0.3cm 1.7cm 1.6cm 3.4cm},clip,width=\linewidth]{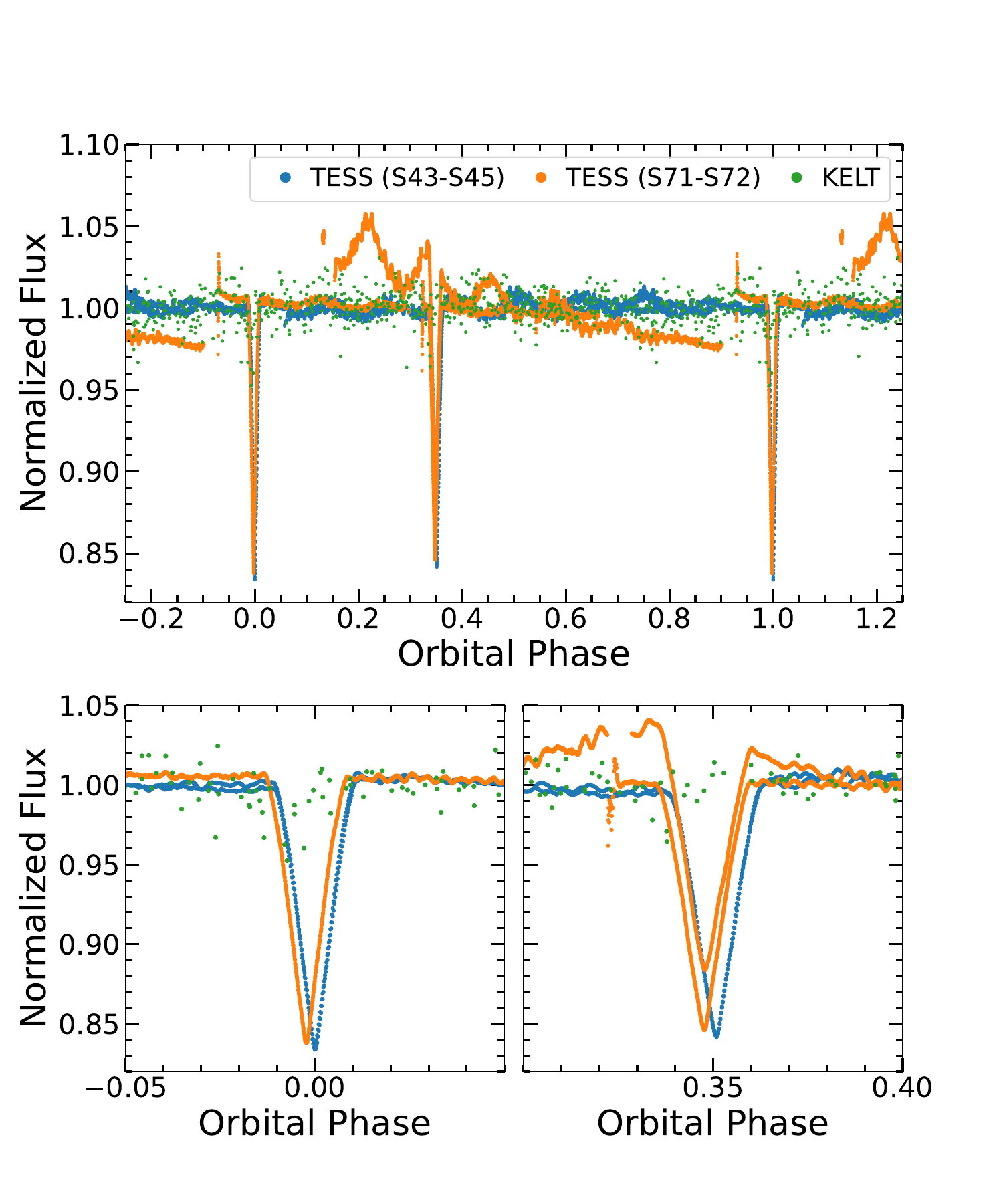}
    \caption{Normalized light curves from TESS (sectors S43-S45 and S71-S72) and KELT surveys in orbital phase. The top panel shows the full orbital phase, and the bottom panels display zoomed-in views of the phase ranges $-0.05$ to $0.05$ (left) and $0.30$ to $0.40$ (right). Blue points correspond to TESS data from sectors S43-S45, orange points to TESS sectors S71-S72, and green points to KELT data.}
    \label{fig:tess_kelt}
\end{figure}

TESS observations show that the timing of the eclipses and depth vary across sectors, and no single period aligns both primary and secondary eclipses consistently, strongly suggesting apsidal motion in the inner binary. Moreover, the eclipses are noticeably shallower in the KELT data (long-term trends have been removed) compared to TESS, even though the stellar components' physical properties should be unchanged (see Fig.\,\ref{fig:tess_kelt}). This points to a variation in orbital inclination, possibly driven by Kozai-Lidov cycles \citep{naoz2016eccentric} induced by the outer Be star. Changes in eccentricity and periastron argument cannot be ruled out either.

Photometric variability seen in KELT outside of eclipses is consistent with density changes in the Be disk at moderate inclination, typical for classical Be stars. These trends further support the scenario of an active and evolving circumstellar environment.

\subsection{Frequency Analysis}

To conduct a frequency analysis of the photometric data of V1371\,Tau, it was necessary to isolate the intrinsic variability of the stellar atmosphere by removing the modulation caused by the eclipses. This was achieved by incorporating the constant flux contribution from the third component into the synthetic light curve of the inner eclipsing binary, and subsequently subtracting the resulting model from the observed light curve. For this analysis, we considered that approximately two-thirds of the total system flux originates from subsystem A, while the remaining one-third is attributed to component B, as estimated in Section\,\ref{sec:V1371Tauspec}.

\begin{figure}
    \centering
    % Answer: [trim={left bottom right top},clip]
    \includegraphics[trim={0.05cm 0.0cm 0 0},clip,width=\linewidth]{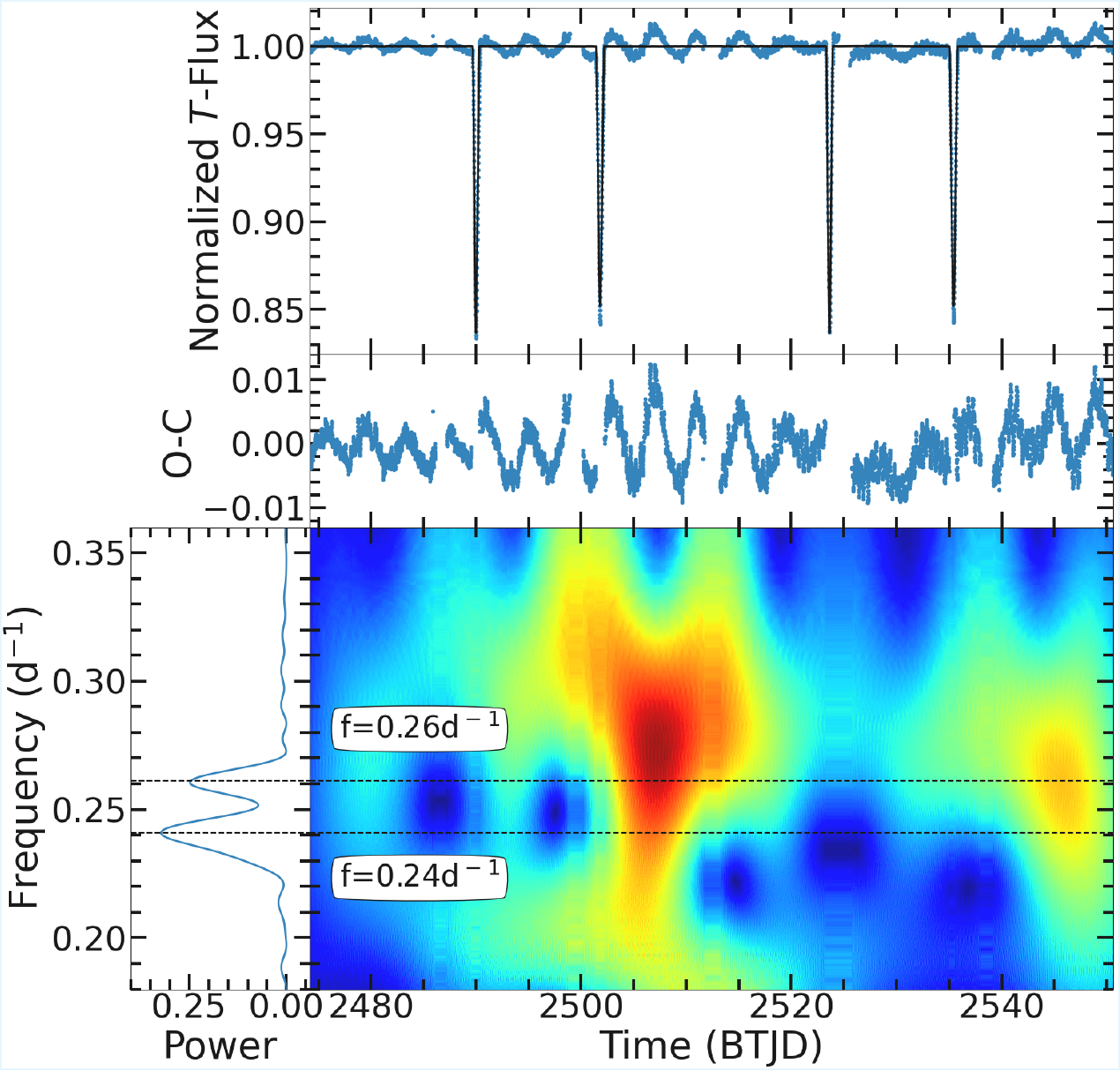}
    \caption{Top panel: Normalized TESS photometric light curve of V1371\,Tau from sectors 43, 44, and 45. Middle panel: Residuals obtained after subtracting the scaled model of the eclipsing binary. Bottom panel: Frequency analysis of the residual light curve of V1371\,Tau using the Lomb-Scargle periodogram (left) and the continuous wavelet transform (right).}
    \label{fig:freq}
\end{figure}

We analysed the residual light curve using Lomb-Scargle periodograms \citep{scargle1982studies, zechmeister2009generalised} and the iterative prewhitening method described by \citet{degroote2009corot} to identify characteristic frequencies, aiming to infer pulsation modes and the rotation period.  In particular, the prewhitening approach fits the function of the original signal in successive steps. The current and previous frequencies are used in a non-linear least-squares fit to the initial light curve as per Eq.\,1 of \citet{burssens2019new}. The process repeats until the S/N exceeds four. The result of the prewhitening is presented in Ap.\,\ref{ap:freq}, and Fig.\,\ref{fig:freq} shows the Lomb-Scargle periodogram and the residual wavelet analysis of the light curve of V1371\,Tau, with the eclipse excess removed to enhance the visualization of the residuals. This excess flux arises due to limitations in the fitting process with \texttt{emcee}, whose optimization does not work well when dealing with light curves with abrupt features, such as narrow eclipses. These features introduce steep gradients in the likelihood surface, making it difficult for the walkers to converge efficiently in parameter space. Despite this, the excess does not affect the physical, orbital or frequency analyses.

This analysis identifies two primary frequencies: $f_1$\,=\,0.24\,d$^{-1}$ and $f_2$\,=\,0.26\,d$^{-1}$ (see Fig.\,\ref{fig:freq}), which lead to a beat frequency of $f_{2-1}$\,=\,0.02\,d$^{-1}$. A similar phenomenon was reported by \citet{emilio2010photometric} in their photometric study of Be star CoRoT-ID 102761769. However, based on the TESS photometric data, this beat frequency corresponds to a beat envelope with a period of approximately 50 days. The stellar origin of these frequencies remains uncertain, but we cannot rule out the possibility that they originate from the star Be. After removing these low-frequency sinusoidal signals, we detect a high-frequency signal at 6.87\,d$^{-1}$. This frequency is characteristic of NRPs and may be associated with the Be star; however, since all components could be pulsators and have similar brightness, this identification remains uncertain.

A third low-frequency peak, $f_3 = 0.226$ c/d, was identified after removing the high-frequency signals. It forms a coherent triplet with the dominant low-frequency peaks at 0.241 and 0.261 c/d. Its high signal-to-noise ratio (11.6) and proximity to the other two peaks suggest that it is physically significant, possibly arising from rotational splitting or binarity-induced modulation.

To perform a seismic diagnostic, we conducted a prewhitening analysis on the residuals of the modelled light curve, aiming to identify characteristic frequencies. However, the analysis revealed insufficient frequencies in the range required to apply the Takata method \citep{takata2020diagnostic}, which depends on detecting a clear pattern of progressively spaced sectoral modes. This lack of sufficient and distinct frequency patterns in the data prevents us from confidently linking the detected frequencies to specific physical mechanisms, such as stellar rotation or g-mode NRPs, to any of the stars within the system.

\section{Conclusion} \label{sec:conclusion}

In this work, we present a study of V1371\,Tau, located at a distance of 1.20$\pm$0.03\,kpc from Earth, based on its variability using TESS photometry and ground-based spectroscopy data, which allowed us to obtain some orbital and physical characterizations of this rare system. Our spectral data from multiple epochs and wavelengths provided insights into the hierarchical nature of the system, concluding that V1371\,Tau is characterized as three main-sequence stars, composed of an eclipsing binary with B-type stars, orbited by a classical Be star, so that we have a complex (B1V\,+\,B0V)\,+\,B0Ve system, with this spectral classification being carried out based on the parameters obtained previously. The identification and characterization of the third body as classical Be was possible thanks to \citet{Rivinius2025} who identified a third object as part of the system with an angular separation of 5\,$mas$, and this work which inferred that the variability of the emission lines in H$\alpha$ does not present a periodicity similar to that of the $\sim$34\,d eclipsing binary.

Modeling the visible spectrum allowed us to deduce atmospheric parameters. At the same time, it was possible to construct the radial velocity curve of the inner eclipsing system, resulting in an inner mass ratio of q$_{\rm A}$=0.78. Applying the merged WDC code and \texttt{emcee} to the TESS light curve data yielded temperatures of 25,000\,K and 28,000\,K for the primary and secondary components, respectively, of the eclipsing system. The eccentricity of the inner binary was measured to be $\sim$0.27 with the argument of periastron, resulting in 150$^\circ$. Finally, values of 14\,M$_\odot$ and 11\,M$_\odot$, and 4.9\,R$_\odot$ and 4.8\,R$_\odot$, were obtained for the masses and radii of the primary and secondary components of the inner eclipsing system. %The characteristics of the outer Be star were studied from its system distance and angular separation. If we assume a circular, coplanar orbit, and the Be star close to maximum elongation, the following orbit could be inferred: $\sim$1,300\,R$_\odot$ for the outer semi-major axis, and assuming a mass of 14\,M$_\odot$ for the Be star, we consequently infer a measurement of $\sim$2.50\,yrs for the orbital period of the outer Be component. These results for the outer component of V1371\,Tau may change substantially as a high value for the eccentricity of the orbit between the A+B system is found.
Considering the average values of the spectral type of the Be star, together with the measurement of its rotation velocity, we infer that it rotates in a range of 44-76\% of its critical rupture velocity. This result for classical Be stars is within the expected range in the literature \citep{zorec2016critical}.

TESS data revealed greater complexity in the system originally studied with KELT photometry. While primary eclipses should have been detectable in KELT, they were likely obscured by variability from the Be star's disk and orbital changes. TESS observations indicate eclipse timing and depth variations consistent with apsidal motion, and the shallower KELT eclipses suggest changes in orbital inclination, possibly due to Kozai-Lidov cycles and eccentricity. Out-of-eclipse variability in KELT also supports an evolving Be disk.

The residual from modeling the light curve of the inner binary system is characterized by three low- and high-frequency sinusoidal signals. We could not identify which star these low-frequency signals (near 0.24\,c/d) come from, and the signals in the TESS light curve of lower amplitude, but visible due to their higher frequency (near 6.87\,c/d), are possible results of NRP from the outer Be star.

A study of the evolutionary status of the stars of V1371\,Tau was carried out. An age of 10\,Myrs has been inferred for the components of V1371\,Tau, ranking them among the youngest stars known. Analyzing the HR diagram (see Fig.\,\ref{fig:hrdiagram}), we can conclude that all components of the triple system are massive stars and are within the main sequence, with secondary eclipsing star being overluminous.

Further improvements to our results can be achieved by further spectral monitoring to disentangle the spectrum of V1371\,Tau and determine individual temperatures, $\log g$ and $v \sin i$, as well as by accurately determining the mass ratio of the outer orbit, incorporating the construction of the radial velocity curve of the system. Furthermore, improved precision in the individual measurements of the extinction factor of the components of V1371\,Tau will contribute to refining our findings. New interferometric data will be crucial to construct the inner and outer orbits and thus complete the characterization of all system components. We did not find X-ray data from the Swift, XMM-Newton, and NuSTAR observatories, which are important to complement studies on the variability of the Be star.

Ultimately, this study provides a comprehensive foundation for future investigations into the intricate dynamics and variable characteristics of V1371\,Tau, contributing valuable insights to the broader understanding of massive stars, hierarchical systems, and classical Be stars.

\begin{acknowledgments}
    % Danilo
    DFR is thankful for the support of the CAPES and FAPERJ/DSC-10 (SEI-260003/001630/2023).

    %Marcelo
    M.E. gratefully acknowledges the financial support of the ``Fen\^omenos Extremos do Universo" of the Funda\c{c}\~ao Arauc\'aria grant 348/2024 and by the State of Paraná Secretary of Science, Technology and Higher Education - Fundo Paraná grant 031/2024.

    % Romualdo
    R.E. thankful for supported by CAPES (Finance Code 001: 88887.914708/2023-00).

    % labadie-bartz
    JLB acknowledges support from the European Union (ERC, MAGNIFY, Project 101126182). Views and opinions expressed are, however, those of the authors only and do not necessarily reflect those of the European Union or the European Research Council. Neither the European Union nor the granting authority can be held responsible for them.

    % SOAR
    Based in part on observations obtained at the Southern Astrophysical Research (SOAR) telescope, which is a joint project of the Minist\'{e}rio da Ci\^{e}ncia, Tecnologia e Inova\c{c}\~{o}es (MCTI/LNA) do Brasil, the US National Science Foundation’s NOIRLab, the University of North Carolina at Chapel Hill (UNC), and Michigan State University (MSU). This research made use of Astropy, a community-developed core Python package for Astronomy (Astropy Collaboration, 2013, 2018).

    % BeSS
    This work has made use of the BeSS database, operated at LIRA, Observatoire de Meudon, France: http://basebe.obspm.fr. In particular, we would like to thank the observers who shared their work on V1371\,Tau, making this manuscript more faithful: Joan G. Fló, Christian Buil, Olivier Garde and James R. Foster.

    % HERMES
    Based on observations obtained with the HERMES spectrograph, which is supported by the Research Foundation - Flanders (FWO), Belgium, the Research Council of KU Leuven, Belgium, the Fonds National de la Recherche Scientifique (F.R.S.-FNRS), Belgium, the Royal Observatory of Belgium, the Observatoire de Genève, Switzerland and the Thüringer Landessternwarte Tautenburg, Germany. 

    % Mercator
    Based on observations made with the Mercator Telescope, operated on the island of La Palma by the Flemish Community, at the Spanish Observatorio del Roque de los Muchachos of the Instituto de Astrofísica de Canarias.
    
\end{acknowledgments}

\begin{contribution}
%%This section gives authors the space to recognize author contributions. The text inside this environment is NOT counted towards the total word quanta. At a minimum, manuscripts are expected to include this text:

DFR identified and determined the physical and orbital parameters of the object, in addition to performing frequency analysis and evolutionary study. ML determined the rotation speed and supervised all the methodology used and the results obtained. JLB and CN were responsible for characterizing the external component as a Be star, in addition to helping define the methodology and obtain the results. JB, TS, LA, MAM and FN were responsible for obtaining the spectroscopic data. AM, EJP, RE and AWP assisted in the analysis and characterization of the triple system. All authors was responsible for writing the manuscript.

%All authors contributed equally to this manuscript.

%% But authors are expected to provide more specific details, e.g. 
%%
%%SC was responsible for writing and submitting the manuscript.
%%WWM came up with the initial research concept and edited the manuscript.
%%OTS obtained the funding and edited the manuscript.
%%EBF provided the formal analysis and validation. He also edited the manuscript.
%%GEH Supervised the undergraduates, wrote the software and administers the project github and Zenodo repositories.
%%
%% Authors can use the Contributor Role Taxonomy (CRediT) at
%% for ideas on how write a good statement tailored to their needs.

\end{contribution}

\facilities{TESS, SOAR, HERMES(Mercator)}

\software{\texttt{astropy} \citep{2013A&A...558A..33A,2018AJ....156..123A,2022ApJ...935..167A}, \texttt{Lightkurve} \citep{collaboration2018lightkurve},
Wilson \& Devinney code \citep{wilson1971realization} and
\texttt{emcee} \citep{foreman2013emcee}
}


\appendix

\section{Radial Velocity of the Inner Binary in V1371 Tau}\label{ap:rv}

\begin{figure*}
    \centering
    % Answer: [trim={left bottom right top},clip]
    \includegraphics[width=0.98\linewidth]{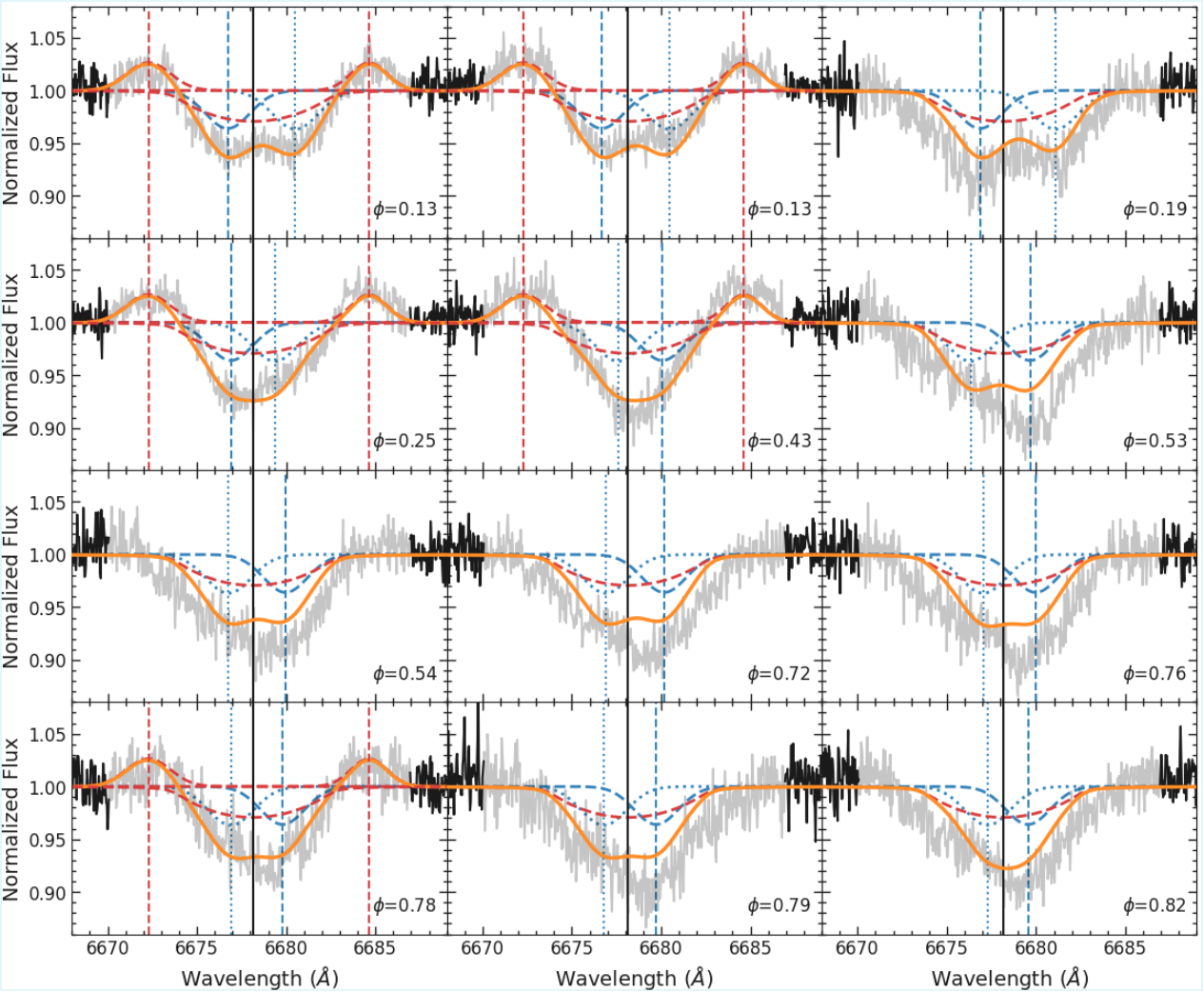}
    \caption{Fits of the HeI\,$\lambda$\,6678\,{\text{\AA}} spectral lines to the components of V1371\,Tau obtained with the HERMES spectrograph. The blue dashed and dotted lines represent the spectral lines of the a and b stars, respectively, the red dashed lines are the fits of the wings in emission and the absorption contribution from the Be star, the orange solid lines are the best fits of the total contribution of the components, and the black solid lines the observed spectrum, with the baseline highlighted. The vertical lines represent the laboratory wavelength (in black) and the wavelength of the center of each Gaussian (in blue and red).}
    \label{fig:rv6678}
\end{figure*}

The heliocentric radial velocity measurements of the inner components of V1371\,Tau, obtained from HERMES spectra at different orbital phases using the HeI\,$\lambda$4922 and HeI\,$\lambda$6678 lines, are presented in Table\,\ref{tab:rv}. Figure\,\ref{fig:rv6678} illustrates the Gaussian fits to the HeI\,$\lambda$6678 line, highlighting the decomposition into the two stellar components, the Be-star absorption and emission wings, and the resulting combined model compared with the observed spectrum.

\begin{table*}[ht]
\caption{Radial velocity measurements of the inner components of V1371,Tau, corrected by the heliocentric reference frame, derived from HERMES spectra.}
\label{tab:rv}      % is used to refer this table in the text
\centering                      % used for centering table
\begin{tabular}{c c c c c c}        % centered columns (4 columns)
\hline\hline
MTO$^*$ & \multicolumn{1}{l}{\multirow{2}{*}{Orbital Phase}} & Comp.\,a & Comp.\,b & Comp.\,a & Comp.\,b \\ % table heading
(BTJD) & & (km\,s$^{-1}$) & (km\,s$^{-1}$) & (km\,s$^{-1}$) & (km\,s$^{-1}$) \\
\hline

&& \multicolumn{2}{c}{$\lambda_0$ = 4921.931$\text{\AA}$} & \multicolumn{2}{c}{$\lambda_0$ = 6678.158$\text{\AA}$} \\
    3,267.755 & 0.133 & -55.40 $\pm$ 13.37 & 149.69 $\pm$ 12.29 & -49.63 $\pm$ 13.51 & 118.76 $\pm$ 14.21 \\
    2,830.749 & 0.135 & -37.54 $\pm$ 11.02 & 162.51 $\pm$ 11.29 & -35.59 $\pm$ 13.52 & 131.39 $\pm$ 17.08 \\
    2,664.453 & 0.189 & -100.75 $\pm$ 11.91 & 95.51 $\pm$ 13.57 & -85.52 $\pm$ 4.28 & 99.70 $\pm$ 12.54 \\
    2,834.735 & 0.253 & -24.70 $\pm$ 10.19 & 143.88 $\pm$ 13.95 & -25.64 $\pm$ 10.39 & 82.61 $\pm$ 13.94 \\
    3,042.405 & 0.430 & 39.43 $\pm$ 9.38 & -60.75 $\pm$ 11.03 & 36.16 $\pm$ 4.36 & -41.51 $\pm$ 11.64 \\
    2,171.778 & 0.535 & 79.50 $\pm$ 8.17 & -91.60 $\pm$ 4.74 & 68.03 $\pm$ 4.91 & -55.96 $\pm$ 14.42 \\
    2,205.613 & 0.541 & 85.45 $\pm$ 9.85 & -61.00 $\pm$ 13.63 & 66.35 $\pm$ 8.44 & -58.52 $\pm$ 20.40 \\
    2,682.404 & 0.723 & 59.83 $\pm$ 9.00 & -89.49 $\pm$ 8.17 & 62.71 $\pm$ 6.53 & -80.53 $\pm$ 10.56 \\
    2,246.548 & 0.759 & 79.08 $\pm$ 10.17 & -68.44 $\pm$ 15.33 & 55.93 $\pm$ 11.25 & -68.95 $\pm$ 21.16 \\
    3,020.397 & 0.776 & 56.97 $\pm$ 11.07 & -73.93 $\pm$ 12.52 & 43.50 $\pm$ 12.59 & -84.33 $\pm$ 9.71 \\
    2,953.522 & 0.787 & 84.84 $\pm$ 10.73 & -75.75 $\pm$ 6.78 & 56.77 $\pm$ 10.91 & -72.17 $\pm$ 12.41 \\
    2,248.479 & 0.816 & 68.15 $\pm$ 12.37 & -72.40 $\pm$ 17.17 & 37.50 $\pm$ 6.96 & -58.70 $\pm$ 26.36 \\

\hline \hline
\raggedright
$^*$Mid-Time Observation.
\end{tabular} \\
\end{table*}

%%%%%%%%%%%%%%%%%%%%%%%%%%%%%%%%%%%%%%%%%%%%%%%%%%%%%

\section{Frequencies Analysis of the V1371 Tau}\label{ap:freq}

\begin{figure}%[!ht]
    \footnotesize
    \centering
    % Answer: [trim={left bottom right top},clip]
    \includegraphics[trim={0 0 0 0},clip, width=\linewidth]{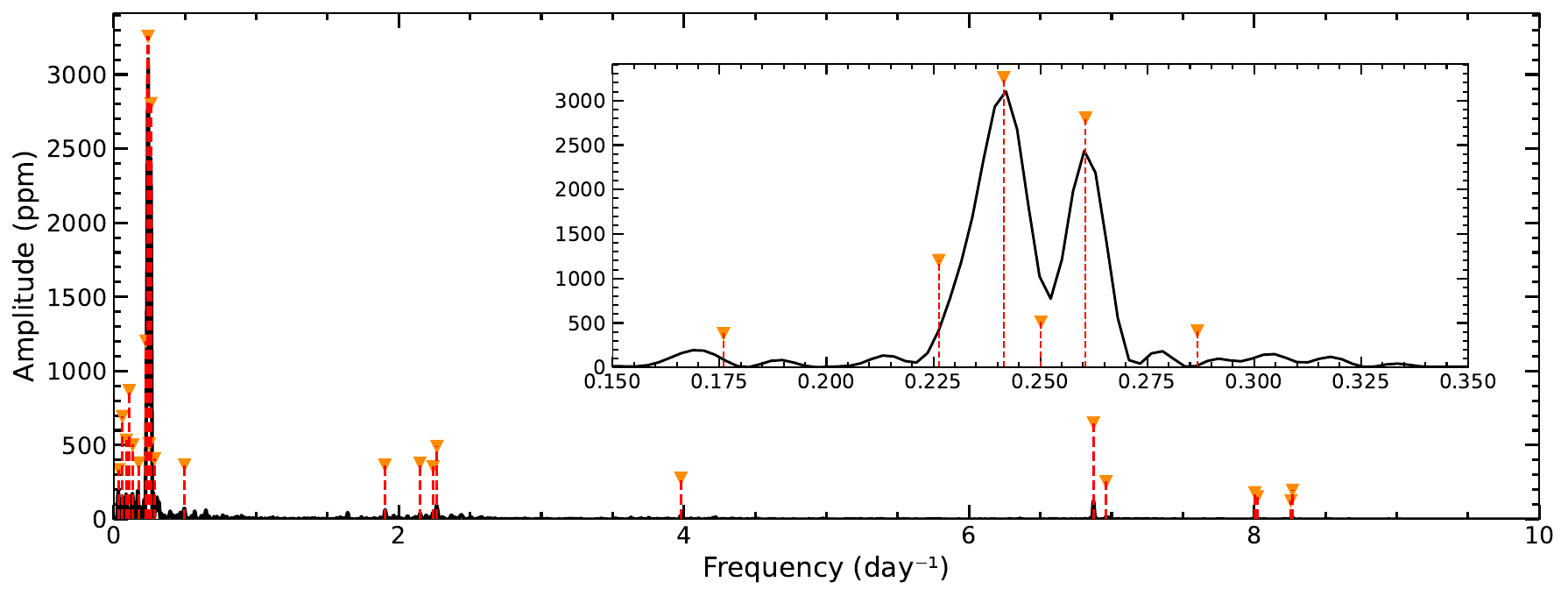}
    \caption{Frequency spectrum obtained with the prewhitening method for the residual of the eclipsing binary fit for the star V1371\,Tau. The main panel shows the full spectrum up to 10\,day$^{-1}$, while the inset highlights the region between 0.05 and 0.3\,day$^{-1}$, emphasizing the detected frequencies (orange symbols).}
    \label{fig:prewspec}
\end{figure}

Independent frequencies were extracted from the residual light curve of the eclipsing binary V1371\,Tau using an iterative prewhitening method. The selected frequencies, listed in Table\,\ref{tab:freq} and illustrated in Figure\,\ref{fig:prewspec}, lie within a 20\,c/d window and have S/N\,$\geq$\,4.0. Among them, the most significant peaks are found at 0.24 and 0.26\,c/d, as well as at 6.87\,c/d, which exhibit particularly high amplitudes and signal-to-noise ratios.

\begin{table}
\caption{Frequencies found from the residual of the eclipsing binary fit to the light curve for the star V1371\,Tau, with signal-to-noise ratio greater than or equal to 4.0 and a frequency window of 20\,c/d.} 
\label{tab:freq}      % is used to refer this table in the text
\centering                      % used for centering table
\begin{tabular}{c c c c}        % centered columns (4 columns)
\hline\hline
Freq (c/d) & Amplitude (ppm) & S/N \\
\hline
0.038 & 320.32 & 4.411 \\
0.062 & 662.59 & 7.579 \\
0.089 & 510.61 & 5.617 \\
0.109 & 827.81 & 8.446 \\
0.133 & 479.27 & 5.840 \\
0.176 & 364.21 & 4.884 \\
0.226 & 1146.72 & 11.649 \\
0.241 & 3103.45 & 24.126 \\
0.250 & 486.36 & 4.407 \\
0.261 & 2671.98 & 23.441 \\
0.287 & 391.36 & 4.511 \\
0.498 & 350.89 & 4.514 \\
1.902 & 348.85 & 4.529 \\
2.148 & 362.41 & 4.421 \\
2.241 & 338.06 & 4.274 \\
2.268 & 469.42 & 5.383 \\
3.979 & 265.26 & 4.020 \\
6.873 & 619.22 & 14.039 \\
6.961 & 242.63 & 6.359 \\
8.004 & 172.54 & 5.805 \\
8.023 & 145.89 & 4.584 \\
8.258 & 120.44 & 4.466 \\
8.269 & 187.50 & 6.398 \\
11.717 & 60.09 & 4.512 \\
14.875 & 36.38 & 4.073 \\
15.136 & 43.55 & 4.449 \\
\hline \hline \\
\end{tabular} \\
\end{table}

%%%%%%%%%%%%%%%%%%%%%%%%%%%%%%%%%%%%%%%%%%%%%%%%%%%%%

%% For this sample we use BibTeX plus aasjournalv7.bst to generate the

\clearpage

\bibliography{ref}{}
\bibliographystyle{aasjournalv7}



\end{document}